\documentclass{aa}  

\bibpunct{(}{)}{;}{a}{}{,} % to follow the A&A style
\usepackage{graphicx}
\usepackage{txfonts}
\newcommand{\Sec}[1]{\S\ref{#1}}
\newcommand{\Sub}[1]{\S\S\ref{#1}}

\newcommand{\Eq}[1]{equation~(\ref{#1})}
\newcommand{\Fig}[1]{Fig.~\ref{#1}}

\usepackage[normalem]{ulem}
\usepackage{xcolor}

\begin{document} 

    \titlerunning{Characterizing structure formation through instance segmentation}
    \authorrunning{D. López-Cano et al.}

   \title{Characterizing structure formation through instance segmentation}

   %\subtitle{}

   \author{
        Daniel L\'{o}pez-Cano
        \inst{1,2}\fnmsep\thanks{daniellopezcano13@gmail.com},
        Jens Stücker
        \inst{1},
        Marcos Pellejero Ibañez
        \inst{3},
        Ra\'{u}l E. Angulo
        \inst{1,4},
        \and
        Daniel Franco-Barranco
        \inst{1,5}
    }

   \institute{
        Donostia International Physics Center (DIPC), Paseo Manuel de Lardizabal, 4, 20018 Donostia-San Sebastián, Spain.
        %\email{c.ptolemy@hipparch.uheaven.space}
        \and
        Departamento de F\'{\i}sica Te\'{o}rica, M\'{o}dulo 15, Facultad de Ciencias, Universidad Aut\'{o}noma de Madrid (UAM), 28049 Madrid, Spain.
        \and
        Institute for Astronomy, University of Edinburgh, Royal Observatory, Blackford Hill, Edinburgh, EH9 3HJ, UK.
        \and
        IKERBASQUE, Basque Foundation for Science, 48013, Bilbao, Spain.
        \and
        Department of Computer Science and Artificial Intelligence, University of the Basque Country (UPV/EHU), Donostia-San Sebastián, Spain.
    }

   \date{Received September XXX; accepted YYY}

% \abstract{}{}{}{}{} 
% 5 {} token are mandatory
 
  \abstract
    {Dark matter haloes form from small perturbations to the almost homogeneous density field of the early universe. Although it is known how large these initial perturbations must be to form haloes, it is rather poorly understood how to predict which particles will end up belonging to which halo. However, it is this process that determines the Lagrangian shape of proto-haloes and is therefore essential to understand their mass, spin and formation history. We present a machine learning framework to learn how the proto-halo regions of different haloes emerge from the initial density field. We develop one neural network to distinguish semantically which particles become part of \emph{any} halo and a second neural network that groups these particles by halo membership into different instances. This instance segmentation is done through the Weinberger method, in which the network maps particles into a pseudo-space representation where different instances can be distinguished easily through a simple clustering algorithm. Our model reliably predicts the masses and Lagrangian shapes of haloes object-by-object, as well as other properties like the halo-mass function. We find that our model extracts information close to optimal by comparing it to the degree of agreement between two N-body simulations with slight differences in their initial conditions. We publish our model open-source and suggest that it can be used to inform analytical methods of structure formation by studying the effect of systematic manipulations of the initial conditions.}


   \maketitle
%
%-------------------------------------------------------------------

\section{Introduction}\label{sec_Introduction} 

Dark matter (DM) haloes are the primary structures in the universe within which galaxies form and evolve. Acting as gravitational anchors, they play a pivotal role in connecting theoretical cosmology with empirical observations from galaxy surveys. Given their significance in cosmology, a comprehensive understanding of DM haloes and their behaviour is paramount. Currently, our most detailed insights into their formation and properties come from N-body simulations \citep[see][for a review]{2012AnP...524..507F}. These computationally intensive simulations model the interactions of vast numbers of particles, pinpointing the regions of the density field where gravitational collapse leads to the formation of DM haloes \citep[e.g.][]{2022LRCA....8....1A}. Therefore, understanding the formation and behaviour of DM haloes is essential to bridge the gap between theoretical models and observational data.

However, providing quick and accurate predictions (based on the initial conditions of a simulation) remains a challenging task for physically-motivated models. An accurate model for halo formation must be able to capture the nonlinear growth of density fluctuations. Previous analytical or semi-analytical models for halo formation, such as the top-hat spherical collapse~\citep{1972ApJ...176....1G, 1977ApJ...218..592G, 1980lssu.book.....P}, the Press-Schechter / Excursion Set Theory~\citep{1974ApJ...187..425P, 1991ApJ...379..440B, 1993MNRAS.262..627L}, or ellipsoidal collapse approaches~\citep[e.g.][]{2001MNRAS.323....1S, 2002MNRAS.329...61S}, qualitatively reproduce the behaviour of the halo-mass function and the merging rate of haloes, however, they fail on predicting these quantities accurately~\citep[e.g.][]{2014MNRAS.440..193J}. Further, N-body simulations show the formation of ``peak-less'' haloes, that cannot be accounted for by any of these methods \citep{2011MNRAS.413.1961L}.

Traditional analytical methods have provided foundational insights into the process of halo formation, but they struggle to capture the full complexity of it. Machine Learning (ML) techniques have emerged as a promising alternative, capable of capturing intricate non-linear dynamics inherent to the gravitational collapse of structures. 
ML algorithms can be trained on N-body simulations to emulate the results of much more expensive calculations. Previous studies have trained ML models to map initial positions and velocities of particles to their final states~\citep{2019PNAS..11613825H, 2019arXiv191004255G, 2020arXiv201200240A, 2021ApJ...913....2W, 2022arXiv220604573J} and to predict the distribution of non-linear density fields \citep{2018ComAC...5....4R, 2019ComAC...6....5P, 2021arXiv211106393S, 2023arXiv230512222Z, 2023arXiv231006929S}.

Further, ML has been used to predict and gain insights into the formation of haloes. Some studies utilized classification methods to anticipate if a particle will become part of a halo~\citep{2018MNRAS.479.3405L, 2022A&C....3800527C, 2023arXiv230502122B}, or to predict its final mass category~\citep{2019MNRAS.490..331L}. In~\cite{2020arXiv201110577L} a regressor network is trained to predict the final halo mass for the central particle in a given simulation crop. The work by~\cite{2020MNRAS.496.5116B} demonstrates how ML-segmentation techniques can be applied to predict halo Lagrangian regions. In~\cite{2019MNRAS.482.2861B} a semantic segmentation network is trained to predict Peak-Patch-haloes. In \cite{2023MNRAS.524.1746L} a network is trained to predict the mass of haloes when provided with a Lagrangian region centred on the centre-of-mass of proto-halo patches and is then used to study assembly bias when exposed to systematic modifications of the initial conditions.

While interesting qualitative insights have been obtained in these studies, it would be desirable to develop a model that accurately predicts halo membership at a particle level, surpassing some of the limitations from previous works. An effective model should predict particles forming realistic N-body halos, improving upon previous models restricted to simpler halo definitions \citep[e.g.][where Peak-patch haloes are targeted]{2019MNRAS.482.2861B}. Additionally, an ideal model should be able to predict disconnected Lagrangian halo patches, overcoming the limitations of methods like the watershed technique used in~\cite{2020MNRAS.496.5116B}, which can only handle simply connected regions. Furthermore, particles within the same halo should share consistent mass predictions, avoiding having different halo mass estimates for particles belonging to the same halo.

We present a general ML framework to predict the formation of haloes from the initial linear fields. We create a ML model designed to forecast the assignment of individual particles from the initial conditions of an N-body simulation to their respective haloes. To do so we train two distinct networks, one for conducting semantic segmentation and another for instance segmentation. These two networks together conform what is known as a panoptic-segmentation model. Our model effectively captures the dynamics of halo formation and offers accurate predictions. We provide the models used in this study for public access through our GitHub repository: \url{https://github.com/daniellopezcano/instance_halos}.

The rest of this paper is organized as follows: In \Sec{sec_Methodology}, we define the problem of identifying different Lagrangian halo regions from the initial density field (\Sub{subsec_21_Phrasing}), introduce the panoptic segmentation method (\Sub{subsec_23_panoptic_segmentation}), present the loss function employed to perform instance segmentation (\Sub{subsec_24_weinberger_loss}), describe the simulations used for model training (\Sub{subsec_25_dataset}), asses the level of indetermination for the formation of proto-haloes (\Sub{subsec_22_Chaos}), outline the CNN architecture (\Sub{subsec_26_Vnet}), and explain our training process (\Sub{subsec_27_training}). In \Sec{sec_Results}, we present the outputs of our semantic model (\Sub{subsec_31Semantic}) and our instance segmentation approach (\Sub{subsec_32Instance}). We investigate how our model reacts to changes in the initial conditions in \Sub{subsec_41_exp_dens} \& \Sub{subsec_42_exp_tidal}, and study how the predictions of our model are affected when varying the cosmology \Sub{subsec_43_exp_s8}. We conclude with a summary and final thoughts in \Sec{sec_Conclusions}.

\section{Methodology}\label{sec_Methodology}

We aim to predict the formation of DM haloes provided an initial density field. To comprehensively address this problem, we divide this section into distinct parts. In \Sub{subsec_21_Phrasing}, we explain the problem of predicting halo-collapse and discuss the most general way to phrase it. In \Sub{subsec_23_panoptic_segmentation}, we introduce the panoptic segmentation techniques and explain how they can be employed to predict halo formation. We divide \Sub{subsec_23_panoptic_segmentation} into two separate parts: semantic segmentation and instance segmentation. In \Sub{subsec_24_weinberger_loss} we describe the loss function employed to perform instance segmentation. In \Sub{subsec_25_dataset}, we present the suite of simulations generated to train and test our models. In \Sub{subsec_22_Chaos} we assess the level of indetermination of proto-halo formation. In \Sub{subsec_26_Vnet} we explain how to build a high-performance model employing convolutional neural networks. Finally, in \Sub{subsec_27_training} we present the technical procedure followed to train our models.

\subsection{Predicting structure formation}\label{subsec_21_Phrasing}
The goal of this work is to develop a machine-learning framework to predict the formation of haloes from the initial conditions of a given universe. Different approaches are possible to define this question in a concrete input/output setting. We want to define the problem in a way that is as general as possible so that our model can be used in many different contexts.

\begin{figure}
  \resizebox{\hsize}{!}{\includegraphics{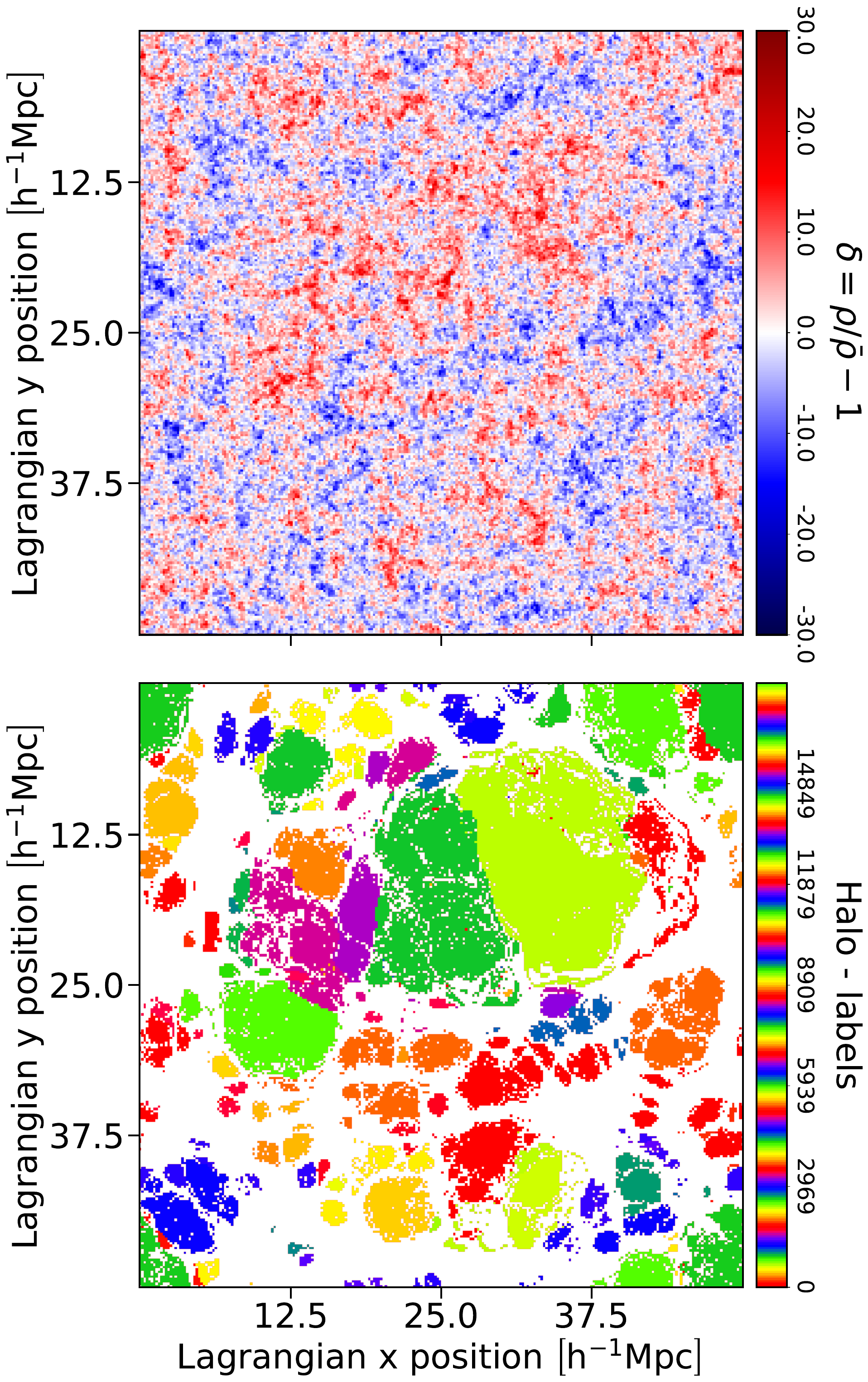}}
  \caption{Example of the prediction problem considered in this article. \textbf{Top panel}: Slice of the three-dimensional initial density field of an N-body simulation. Each voxel (represented here as a pixel) corresponds to a particle that can become part of a halo at later times. \textbf{Bottom panel}: Regions in the initial condition space (same slice as the top panel) that are part of different DM haloes at redshift $z=0$. Pixels coloured in white do not belong to any halo. Pixels with different colours belong to different haloes}. In this work, we present a machine-learning approach to predict the formation of haloes (as in the bottom panel) from the initial condition field (top panel).
  \label{fig_halo_lables}
\end{figure}

The input of the model will be the linear density field discretized to a three-dimensional grid $\delta_{ijk}$. A slice through such a linear density field is shown in the top panel of Figure \ref{fig_halo_lables} and represents how our universe looked in early times, e.g., $z \gtrsim 100$. Beyond the density field, we also provide the linear potential field $\phi_{ijk}$ as an input. The information included in the potential is in principle degenerate with the density field if the full universe is specified. However, if only a small region is provided, then the potential contains additional information of e.g. the tidal field sourced by perturbations outside of the region considered.

The model shall predict which patches of the initial density field become part of which haloes at later times. Concretely, we want it to group the $N^3$ initial grid cells (corresponding, e.g., to particles in a simulation) into different sets so that each set contains exactly all particles that end up in the same halo at a later time. Additionally, there has to be one special extra set that contains all remaining particles that do not become part of any halo:
\begin{align}
    \text{Input:\quad}& \delta_{ijk}, \phi_{ijk} \\
    \text{Output:\quad}& \overbrace{\{\mathrm{id}_A, \mathrm{id}_B, ... \}}^{\text{halo 1}}, \overbrace{ \{\mathrm{id}_C, \mathrm{id}_D, ... \} }^{\mathrm{halo 2}}, ..., \overbrace{\{\mathrm{id}_E, \mathrm{id}_F, ... \}}^{\text{outside of haloes}},
\end{align}

This task is called in the ML literature an \emph{instance segmentation} problem. Note that it is different from typical classification problems since (A) the number of sets depends on the considered input and (B) the sets have no specific order. In practice, it is useful to define the different sets by assigning different number-labels to them. For example, one possible set of particles belonging to the same halo can be given the label ``1'', another set the label ``2'', and so forth. These number-labels do not have a quantitative meaning and are permutation invariant, for example, interchanging the label ``1'' with ``2'' yields the same sets. 

We show such labelling of the initial space in the bottom panel of \Fig{fig_halo_lables}. In this case, the labels were inferred by the membership to haloes in an N-body simulation that employs the initial conditions depicted in the top panel of \Fig{fig_halo_lables} (see Sec.~\ref{subsec_25_dataset}). Our goal is to train a model to learn this instance segmentation into halo sets by training it on the output from N-body simulations.

We note that other studies have characterised the halo-formation processes through a slightly different prediction problem. For example, \cite{2020arXiv201110577L} trains a neural network to predict the final halo masses directly at the voxel level. 
While their approach offers insights into halo formation, our method provides a broader perspective: halo masses can be inferred easily through the size of the corresponding sets, but other properties can be inferred as well -- for example the Lagrangian shapes of haloes which are important to determine their spin \citep{white_1984}. Furthermore, our approach ensures the physical constraint that particles that become part of the same halo are assigned the same halo mass.

\subsection{Panoptic Segmentation}\label{subsec_23_panoptic_segmentation}

The proposed problem requires first to segment the particles semantically into two different classes (halo or non-halo) and then to classify the particles inside the halo class into several different instances. The combination of such semantic plus instance segmentation is sometimes referred to as \emph{panoptic segmentation}. Several strategies have been proposed to solve such panoptic segmentation problems \citep{2016arXiv161108272K, 2016arXiv161108303B, 2017arXiv170402386A, 2017arXiv170802551D, 2018arXiv180100868K, 2023arXiv230402643K} and they usually operate in two-steps:

\begin{enumerate}
    \item \textbf{Semantic segmentation}: The objective of this task is to predict, for each voxel in our initial conditions (representing a tracer particle in the N-body code), whether it will be part of a DM halo at $z=0$. This task is a classification problem, and we will employ the balanced cross-entropy (BaCE) loss~\citep{2015arXiv150406375X} to tackle it:
    
    \begin{equation}\label{eq_BaCE}
    \mathcal{L}_{\textrm{BaCE}}\left(\mathbf{Y}, \mathbf{\hat{Y}}\right) = -\beta \mathbf{Y} \log  \mathbf{\hat{Y}}-(1-\beta)\left(1-\mathbf{Y}\right) \log (1- \mathbf{\hat{Y}})
    \end{equation}

    Here, $\mathbf{Y}$ represents the ground truth data vector, each entry corresponds to a voxel and is equal to $1$ if the associated particle ends up being part of a halo; otherwise, its value is $0$. $\mathbf{\hat{Y}}$ contains the model predictions, with each entry representing the probability that this particle ends up in a halo. The parameter $\beta$ handles the class imbalance and is calculated as the number of negative samples divided by the total number of samples. We measure $\beta$ using our training simulations (see \Sub{subsec_25_dataset}) and obtain a value of $\beta = 0.5815$\footnote{The value of $\beta$ depends on many properties such as the cosmological parameters chosen for the simulations, the redshift, or the mass resolution. We would need to retrain our network and recompute the value of $\beta$ to obtain reliable predictions in different scenarios.}. After training our network, we need to choose a semantic threshold to generate the final semantic predictions. This threshold is calibrated to ensure that the fraction of predicted particles belonging to haloes is equal to $1-\beta$, resulting in a value of $0.589$ (refer to Appendix~\ref{sec_A25} for an in-depth explanation).
    
    \item \textbf{Instance segmentation}: The objective of this task is to recognize individual haloes (instances) by identifying which particles (from those that are predicted to be part of a DM halo) belong to the same object and separating them from others.
    
    Instance segmentation tasks are not conventional classification problems and tackle the problems of having a varying number of instances and a permutational-invariant labelling. To our knowledge, there is no straightforward way to phrase the problem of classifying each voxel into a flexible number of permutable sets through a differentiable loss function. Typical approaches train a model to predict a related differentiable loss and then apply a postprocessing step on top of it. Unfortunately, this leads to the loss function not directly reflecting the true objective.
    
    Various approaches have been proposed to tackle this problem~\citep{2016arXiv161108272K, 2016arXiv161108303B, 2017arXiv170402386A, 2017arXiv170802551D, 2018arXiv180100868K, 2023arXiv230402643K}. A popular method is the watershed technique~\citep{2016arXiv161108272K, 2016arXiv161108303B}. This method uses a network to predict semantic segmentation and the borders of different instances \citep{2018arXiv180710097D} and then applies a watershed algorithm to separate different instances in a post-processing step. However, the watershed approach comes with several limitations:
    \begin{itemize}
        \item It cannot handle the identification of disconnected regions belonging to the same instance, a problem known as occlusion.
        \item It is necessary to select appropriate threshold values for the watershed post-processing step to generate the final instance map. These parameters are typically manually chosen to match some particular metric of interest, but might negatively impact the prediction of other properties. For instance, in \cite{2020MNRAS.496.5116B}, they apply the watershed technique to predict Lagrangian halo regions identified with the \texttt{HOP} algorithm \citep{1998ApJ...498..137E}. However, they choose the watershed threshold to reproduce the halo-mass-function, which does not ensure that the Lagrangian halo regions are correctly predicted.
        \item The watershed approach would struggle to identify the borders of Lagrangian halo regions since they are difficult to define. In \Fig{fig_halo_lables} it can be appreciated that the borders of halo regions are very irregular. There also exist points in the ``interior'' of these regions which are ``missing'' and make it particularly complex to define the border of a halo.
    \end{itemize}
     
     Despite all the challenges presented by the watershed approach, in~\Sec{sec_A1}, we apply this method to predict the formation of FoF-haloes and discuss how the border-prediction problem can be addressed. 

    An approach that offers greater flexibility for grouping arbitrarily arranged particles was presented by \cite{2017arXiv170802551D}. We will follow this approach through the remainder of this work. The main idea behind this method, which we will refer to as the ``Weinberger approach''\footnote{The loss function employed by \cite{2017arXiv170802551D} to perform instance segmentation is inspired by a loss function originally proposed by \cite{weinberger_original} in the context of contrastive learning as a triplet-loss function.}, is to train a model to produce a  ``pseudo-space representation'' for all the elements of our input space (i.e., voxels/particles in the initial conditions). An ideal model would map voxels belonging to the same instance close together in the pseudo-space while separating them from voxels belonging to different instances. Consequently, the pseudo-space distribution would consist of distinct clouds of points, each representing a different instance (see \Fig{fig_2}). The postprocessing step required to generate the final instance segmentation in the Weinberger approach is a clustering algorithm which operates on the pseudo-space distributions.

\end{enumerate}

\subsection{Weinberger loss}\label{subsec_24_weinberger_loss}

The Weinberger approach possesses some advantages over other instance segmentation techniques: First of all, the loss function more closely reflects the instance segmentation objective; that is, to classify different instances into a variable number of permutationally invariant sets. Secondly, the approach is more flexible and makes fewer assumptions, for example, it can handle occlusion cases and does not need to assume the existence of well-defined instance borders.

\begin{figure}
  \resizebox{\hsize}{!}{\includegraphics{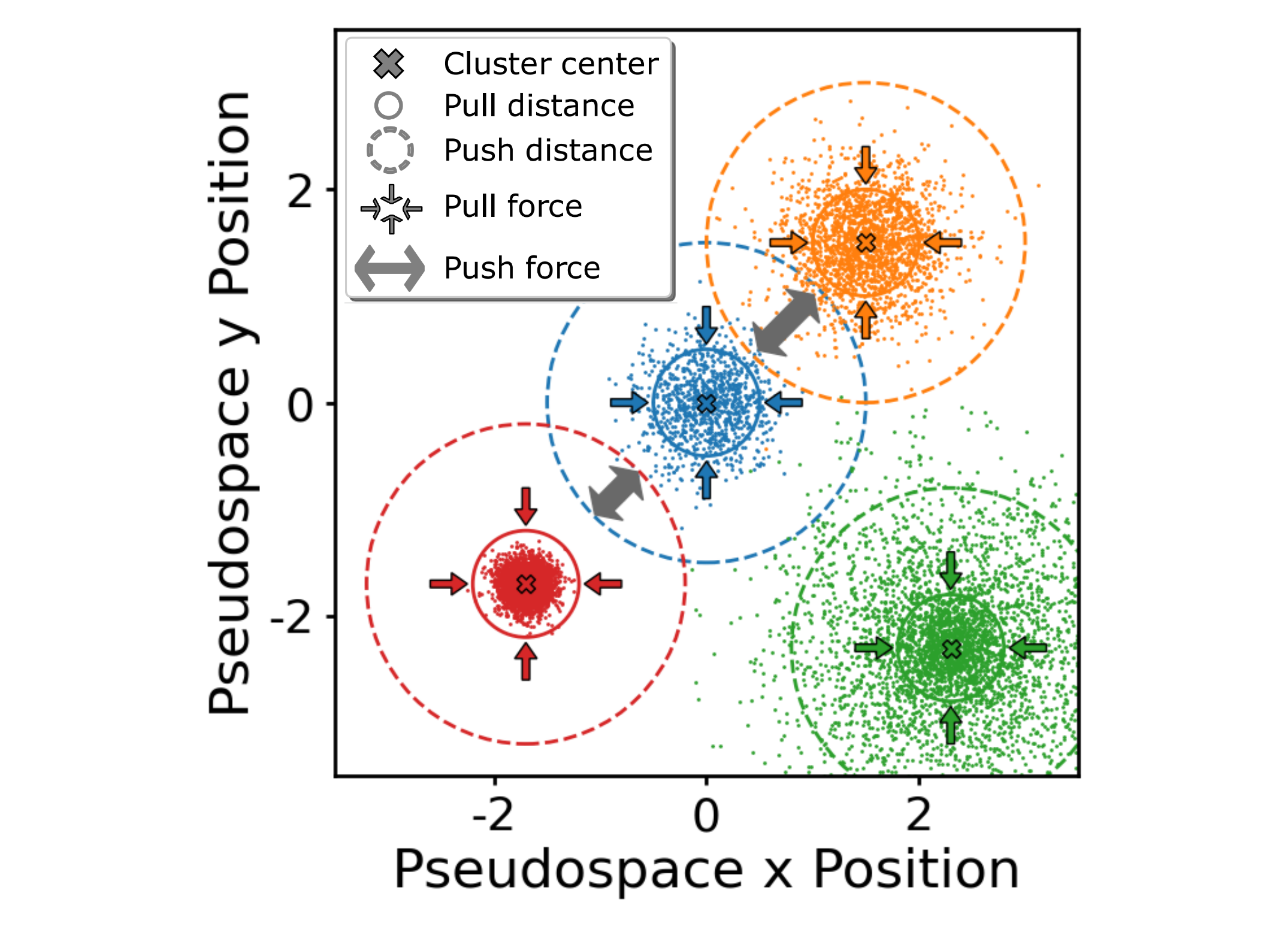}}
  \caption{Example of a two-dimensional pseudo-space employed to separate different instances according to the Weinberger loss. Coloured points represent individual points mapped into the pseudo-space. The centres of the clusters are presented as coloured crosses. Coloured arrows depict the influence of the pull force term, only affecting points outside the $\delta_{\text{Pull}}$ range of their corresponding cluster centre. Grey arrows show the influence of the push force that manifests if two cluster centres are closer than the distance $2\cdot \delta_{\text{Push}}$}
  \label{fig_2}
\end{figure}

In \Fig{fig_2}, we schematically illustrate the effects of the individual components of the Weinberger loss. Each point in this figure represents a pseudo-space embedding of an input voxel. The colours indicate the assigned labels based on the ground truth. Points sharing the same colour belong to the same instance (according to the ground truth), whereas different colours depict separate instances. The "centre of mass" for each cluster is computed and indicated with coloured crosses as "cluster centres". The Weinberger loss is constituted by three separate terms:

\begin{itemize}
    \item \textbf{Pull force,} \Eq{eq_Lpull}:        
    \begin{equation}\label{eq_Lpull}
    L_{pull}=\frac{1}{C} \sum_{c=1}^C \frac{1}{N_c}\sum_{i=1}^{N_c}\max\left( \left(\left\|\mathbf{\mu_c}-\mathbf{x_i}\right\|-\delta_{\text{Pull}}\right)^2 , 0 \right)
    \end{equation}
    Given a certain instance $c$ (where $C$ is the total number of instances), a point $i$ belonging to that set, whose pseudo-space position is $\mathbf{x_i}$, will feel an attraction force proportional to the distance to the instance centre $\mathbf{\mu}_c = \sum_{i=1}^{N_c} \mathbf{x_i} / N_{c}$, where $N_c$ is the number of members associated with the instance $c$. Points closer than $\delta_{\text{Pull}}$ (which is a hyperparameter of the Weinberger loss) from the instance centre will not experience any pull force. The pull force is represented in \Fig{fig_2} as coloured arrows pointing towards the instance centres outside the solid-line circles, which symbolize the distance $\delta_{\text{Pull}}$ to the instance centres.

    \item \textbf{Push force,} \Eq{eq_Lpush}:
    \begin{equation}\label{eq_Lpush}
    L_{\text{push}}=\frac{1}{C(C-1)} \sum_{\substack{c_A=1\\c_A \neq c_B}}^C \sum_{c_B=1}^C \max\left(  \left(2 \delta_{\text{Push}}-\left\|\mathbf{\mu_{c_A}}-\mathbf{\mu_{c_B}}\right\| \right)^2, 0\right)
    \end{equation}
    Two instances $A$ and $B$ will repel each other if the distance between their instance centres in the pseudo-space, $\mu_{c_A}$ and $\mu_{c_B}$, is smaller than $2 \delta_{\text{Push}}$ (a hyperparameter of the Weinberger loss). The force they feel is proportional to the distance between them. In \Fig{fig_2} the push force is represented as grey arrows. The dashed circles represent the distance $\delta_{\text{Push}}$ to the instance centres.

    \item \textbf{Regularization force,} \Eq{eq_Lreg}:        
    \begin{equation}\label{eq_Lreg}
    L_{\text {reg }}=\frac{1}{C} \sum_{c=1}^C\left\|\mathbf{\mu_c}\right\| \\
    \end{equation}
    To avoid having an arbitrarily big pseudo-space distribution all instance centers will feel an attraction towards the pseudo-space origin.
\end{itemize}

The overall effect of these forces on the total Weinberger loss is written as:
\begin{equation}\label{eq_Weinberger}
\mathcal{L}_{\textrm{Wein}}= c_{\text{Pull}} \cdot L_{\text{Pull}} + c_{\text{Push}} \cdot L_{\text{Push}}+ c_{\text{Reg}} \cdot L_{\text{Reg}}
\end{equation}

Where $c_{\text{Pull}}$, $c_{\text{Push}}$, and $c_{\text{Reg}}$ are hyperparameters that regulate the strength of the different components.

Minimizing \Eq{eq_Weinberger} ensures that the pseudo-space mapping produces instance clusters separated from each other. A model trained effectively will predict pseudo-space distributions with points corresponding to the same instances being grouped together and distinctly separated from other instances. In an ideal scenario in which the Weinberger loss is zero, all points are closer than $\delta_{\text{Pull}}$ to their corresponding cluster centres, and clusters are at least $2\delta_{\text{Push}}$ apart. However, realistically, the Weinberger loss won't be exactly zero, necessitating a robust clustering algorithm for accurate instance map predictions.

In Appendix~\ref{sec_A2} we describe the clustering algorithm that we have developed to robustly identify the different instance maps. In our clustering algorithm we first compute the local density for each point in our pseudo-space based on a nearest neighbors calculation. We then identify groups as descending manifolds of density maxima surpassing a specified persistence ratio threshold. Particles are assigned to groups according to proximity and density characteristics. We merge groups selectively, ensuring that the persistence threshold is met. The algorithm relies on three key hyper-parameters for optimal performance: $N_{\rm{dens}}$, $N_{\rm{ngb}}$ and $p_{\rm{thresh}}$. This approach effectively segments the pseudo-space distribution of points, even when perfect separation is not achieved, thus enhancing the reliability of predicted instance maps.

\subsection{Dataset of Simulations}\label{subsec_25_dataset}

We generate twenty N-body simulations with different initial conditions to use as training and validation sets for our panoptic segmentation model. Our simulations are carried out using a lean version of \texttt{L-Gadget3} \citep[see][]{2008MNRAS.391.1685S, 10.1111/j.1365-2966.2012.21830.x, 10.1093/mnras/stab2018}. For each of these simulations, we evolve the DM density field employing ${\rm N_{DM}}=256^3$ DM particles in a volume of $V_{\textrm{box}} = (50\,h^{-1}\mathrm{Mpc})^3$, resulting in a DM particle-mass of $m_{\rm DM}=6.35\cdot10^{8}\,h^{-1}{\rm M_\odot}$. All our simulations employ the same softening length: $\epsilon=5\,h^{-1}\mathrm{kpc}$, and share the cosmological parameters derived by~\cite{2020A&A...641A...6P}, that is, $\sigma_8=0.8288$, $n_{\mathrm{s}} = 0.9611$, $h = 0.6777$, $\Omega_\textrm{b} = 0.048252$, $\Omega_\textrm{m} = 0.307112$, and $\Omega_\Lambda = 0.692888$. Our suite of simulations is similar to the one employed in~\cite{2020arXiv201110577L}.

We use a version of the \texttt{NgenIC} code~\citep{2015ascl.soft02003S} that uses second-order Lagrangian Perturbation Theory (2LPT) to generate the initial conditions at $z=49$. We employ a different random seed for each simulation to sample the Gaussian random field that determines the initial density field. We identify haloes at redshift $z=0$ in our simulations using a Friends-of-Friends algorithm~\citep{1985ApJ...292..371D}, with linking length $b=0.2$. In this work, we will only consider haloes formed by $155$ particles or more, corresponding to $M_\textrm{FoF}	\gtrapprox 10^{11}\,h^{-1}{\rm M_\odot}$. We use $18$ of these simulations to train our model and keep $2$ of them to validate our results.

\subsection{Assessing the level of indetermination}\label{subsec_22_Chaos}

In addition to the training and test sets, we run a set of simulations to establish a target accuracy for our model. These simulations test to what degree small sub-resolution changes of the initial density field can affect the final Lagrangian halo regions.

Structure formation simulations resolve the initial conditions of a considered universe only to a limited degree and exhibit therefore an inherent degree of uncertainty. (1) The numerical precision of simulations is limited (e.g. to 32bit floating point numbers) and therefore any results that depend on the initial conditions beyond machine precision are inherently uncertain. For example, \cite{2019ApJ...871...21G} show that changes in the initial displacement of N-body particles at the machine-precision level can lead to differences in the final locations of particles as large as individual haloes. (2) The initial discretization can only resolve the random perturbations of the Gaussian random field down to a minimum length scale of the mean-particle separation. If the resolution of a simulation is increased, then additional modes enter the resolved regime and act as additional random perturbations. Such additional perturbations may induce some random changes in the halo assignment of much larger-scale structures.

A good model should learn all aspects of structure formation that are certain and well resolved at the considered discretization level. However, there is little use in predicting aspects that are under-specified and may change with resolution levels. Therefore, we conduct an experiment to establish a baseline of how accurate our model shall be.

We run two additional $N=256^3$ simulations with initial conditions generated by \texttt{MUSIC} code~\citep{2011MNRAS.415.2101H}. For these simulations we keep all resolved modes fixed (up to the Nyquist frequency of the $256^3$ grid), but we add to the particles different realisations of perturbations that would be induced by the next higher resolution level. We do this by selecting every $2^3$th particle from two initial condition files with $512^3$ particles and with different seeds at the highest level (``level 9'' in \texttt{MUSIC}). Therefore, the two simulations differ only in the random choice of perturbations that are unresolved at the $256^3$ level. We refer to these two simulations as the ``baseline'' simulations.

\begin{figure}
  \resizebox{\hsize}{!}{\includegraphics{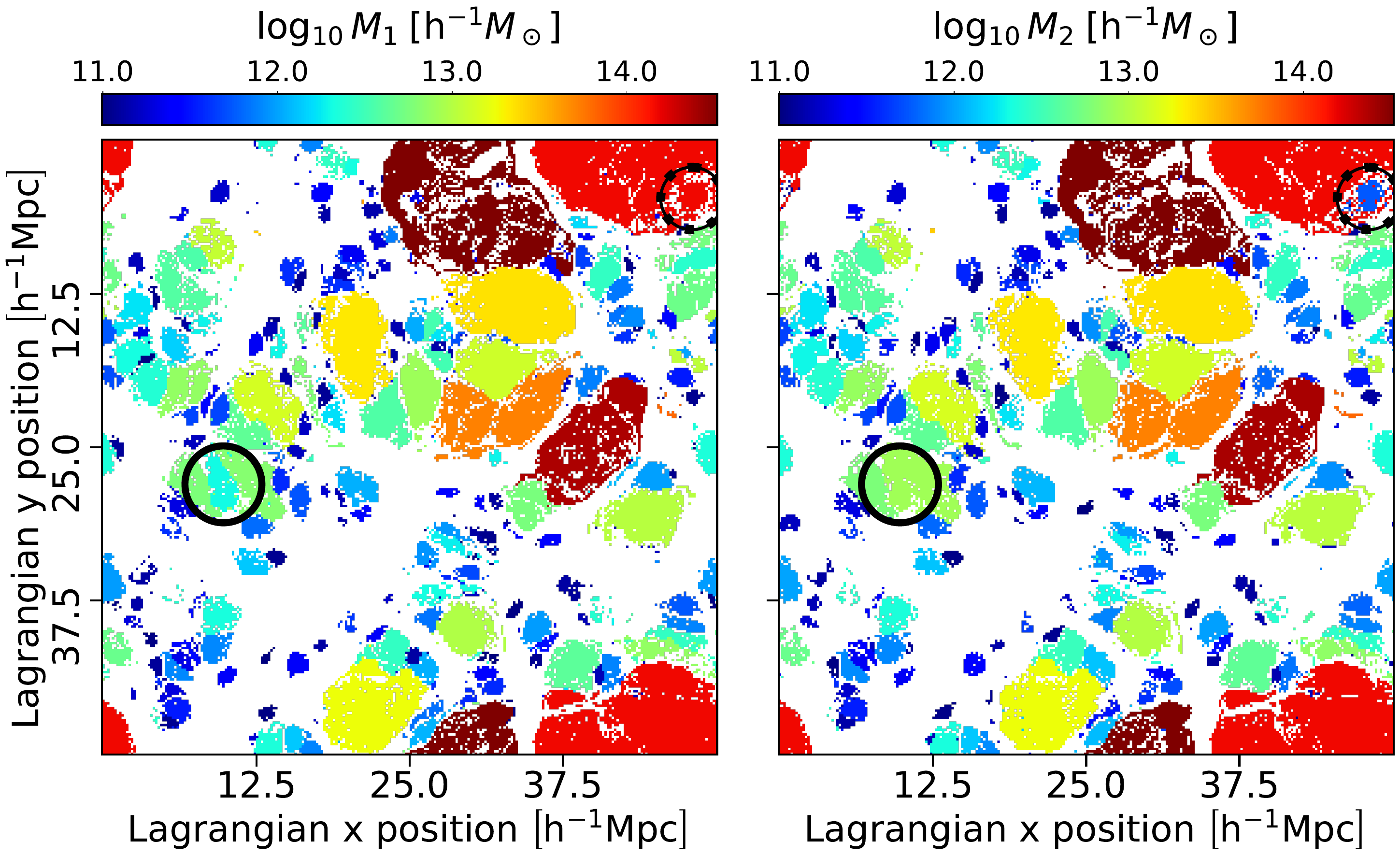}}
  \caption{Slice of the Lagrangian halo regions of the two ``baseline'' simulations (left and right panels respectively). These simulations only differ in sub-resolution perturbations to the initial conditions and their level of agreement sets a baseline for the desired accuracy of our models. The colours employed for both panels represent the mass of the halo associated with each particle for the different Lagrangian halo patches. Circled regions highlight Lagrangian patches whose associated mass significantly changes between the two simulations.}
  \label{fig_1}
\end{figure}

In \Fig{fig_1} we show a slice of the Lagrangian halo patches at $z=0$ through these simulations (left and right panels respectively). The colour map in this Figure represents the masses of the halo that each particle becomes part of, which correspond to the size of the corresponding halo-set. We colour each pixel (which corresponds to a certain particle) according to the mass of the halo that it belongs to. We can appreciate that the outermost regions of the Lagrangian regions are particularly affected while the innermost parts remain unchanged. Notably, in certain instances, significant changes appear due to the merging of haloes in one of the simulations where separate haloes are formed in the other (black-circled regions).

Throughout this article, we will use the degree of correspondence between the baseline simulations as a reference accuracy level. We consider a model close to optimal if the difference between its predictions and the ground truth is similar to the differences observed between the two baseline simulations. A lower accuracy than this would mean that a model has not optimally exploited all the information that is encoded in the initial conditions. A higher accuracy than this level is not desirable, since it is not useful to predict features that depend on unresolved aspects of the simulation and may be changed by increasing the resolution level.

\subsection{V-Net Architecture}\label{subsec_26_Vnet}

\begin{figure}
  \resizebox{\hsize}{!}{\includegraphics{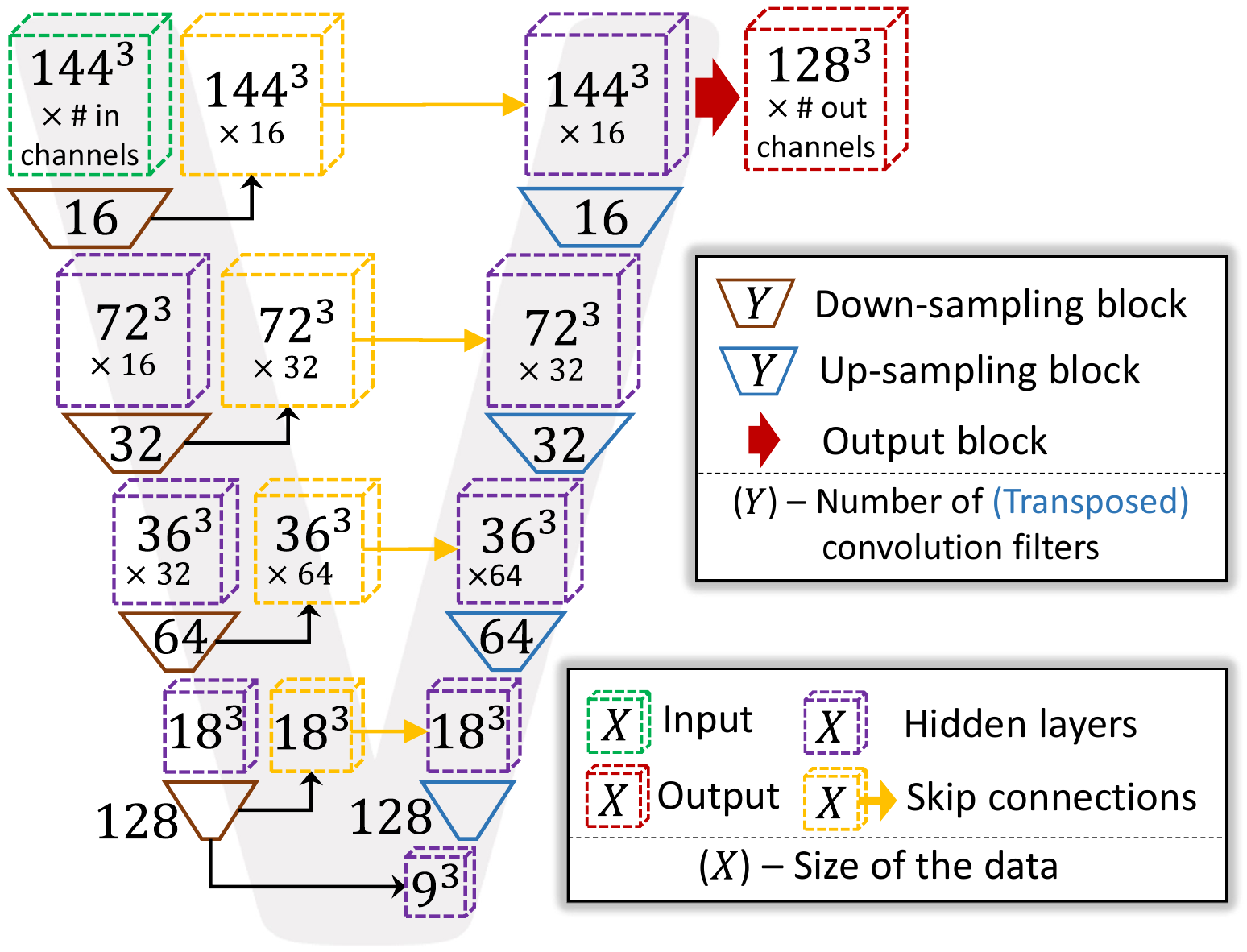}}
  \caption{Flowchart of the particular V-Net architecture we have implemented. The network can take as input multiple channels with dimensions of $144^3$ (top left green cube) and generates predictions for the central voxels with dimensions $128^3$ (top right red cube). The flowchart illustrates the encoder and decoder paths, along with other distinctive features of the network. Notably, the hidden layers and skip connections are represented by purple and yellow cubes, with their respective dimensions annotated at their centres. The down-sampling and up-sampling blocks are shown as brown and purple trapezoids, in their centres we indicate the number of filters employed for the convolution (or transposed convolution) operations.}
  \label{fig_3}
\end{figure}

V-nets are state-of-the-art models, product of many advances in the field of ML over the last decades~\citep{Fukushima1980NeocognitronAS, 726791, NIPS2012_c399862d, 2014arXiv1409.4842S, 2014arXiv1411.4038L, 2015arXiv150504597R, 2015arXiv151203385H}. They are a particular kind of convolutional neural network (CNN) developed and optimized to efficiently map between volumetric inputs and volumetric outputs. V-nets are formed by two separate modules: the encoder (or contracting path) which learns how to extract large-scale abstract features from the input data; and the decoder (or up-sampling path) that translates the information captured by the encoder to voxel-level predictions (also making use of the information retained in the ``skipped connections''). We train V-nets to minimize the loss functions presented in \Sub{subsec_23_panoptic_segmentation} and \Sub{subsec_24_weinberger_loss}. We now explain the technical characteristics of how we have implemented a V-net architecture in \textsc{TensorFlow} \citep{tensorflow2015-whitepaper} (see \Fig{fig_3} for a schematic representation of our network architecture):

\begin{itemize}
    \item Input: Our network is designed to accept as input 3D crops consisting of $144^3$ voxels.\footnote{Ideally, we would prefer to accept as input $256^3$ voxels (corresponding to the full simulation box). However, our GPU resources, though powerful (specifically, an \texttt{NVIDIA QUADRO RTX 8000} with 48 GB of memory), are insufficient to accommodate such an input size while maintaining a reasonably complex network architecture.}  For the results presented in \Sec{sec_Results}, we employ two input channels for the semantic segmentation model, corresponding to the initial density field and the displacement potential, which is defined through Poisson's equation as:
    \begin{align}\label{eq_potential}
        \delta(\vec{q}) &= \vec{\nabla}^2 \phi(\vec{q})
    \end{align}
    For the instance segmentation model, we include three additional input channels corresponding to the Lagrangian positions of particles. This is necessary since the network has to be able to map different haloes with the same density (and potential) structure at different locations in the initial field to different locations in the pseudo space.
    \item Encoder / contractive / down-sampling / down-scaling path: This module consists of consecutive down-scaling blocks that reduce the number of voxels per dimension by half at each level of the network. The purpose of the down-scaling path is to enlarge the network's field of view, enabling per-voxel predictions that take into account distant regions of the field. Achieving this would be impractical using large convolution kernels, as they would consume excessive memory. Within each down-sampling block, we apply three consecutive convolution operations followed by a Leaky-ReLu activation function. The number of convolution filters in a contractive block doubles with each level of compression to improve the performance of the model. For each level, the latent maps computed before the final convolution (the one used to reduce the data size) are temporarily stored to serve as a skip connection for the up-scaling path. In \Fig{fig_3} we show the dimensions of the latent maps computed at each level of the contractive path; the deepest level of our network has a size of $9^3 \times 128$.
    \item Decoder / up-sampling / up-scaling path: This path operates opposite to the contractive path; each up-scaling block doubles the number of voxels per dimension, ultimately recovering an image with the same dimensions as the original input (see \Fig{fig_3}). The up-sampling path facilitates the extraction of smaller-scale features that influence the final per-voxel predictions. Within an up-sampling block, the final convolution is substituted with a transposed convolution operation, that allows doubling the output size per dimension.
    \item Output: The final module of our network takes as input the latent maps with dimensions $144^3 \times 16$. The functionality of this module varies depending on the task at hand. For semantic segmentation, a single convolution operation is performed, resulting in a latent map of $144^3 \times 1$. This map is subsequently cropped to $128^3 \times 1$, and finally, a sigmoid activation function is applied. In the case of instance segmentation, we have decided to work in a three-dimensional pseudo-space, hence, we employ a convolution with three filters to obtain $144^3 \times 3$ maps, which are afterwards cropped to $128^3 \times 3$. In both cases, the final cropping operation is implemented to enhance model performance by focusing on the central region of the image.
\end{itemize}

The V-Net architecture we have implemented is a state-of-the-art model that encompasses over $3\cdot 10^6$ trainable parameters.

\subsection{Training}\label{subsec_27_training}

We train our segmentation networks using a single Nvidia Quadro RTX 8000 GPU card. As mentioned in \Sub{subsec_25_dataset}, we employ $18$ simulations for training, dividing the training process into separate stages for the semantic and instance models.

To ensure robust training and enhance the diversity of training examples without needing to run more computationally expensive simulations, we apply the following data augmentation operations each time we extract a training sample from our simulation suite:
\begin{enumerate}
    \item Select one of the training simulation boxes at random.
    \item Select a random voxel as the center of the input/output regions.
    \item Extract the input ($144^3$) and target ($128^3$) fields of interest by cropping the regions around the central point, considering the periodic boundary conditions of the simulations.
    \item Randomly transpose the order of the three input grid dimensions $q_x, q_y, q_z$.
    \item Randomly chose to flip the axes of the input fields.
\end{enumerate}

To train our semantic and instance segmentation networks we minimize the respective loss functions -- \Eq{eq_BaCE} and \Eq{eq_Weinberger} -- employing the Adam optimizer implemented in \texttt{TensorFlow}~\citep{tensorflow2015-whitepaper}. We train our models for over $80$ epochs, each epoch performs mini-batch gradient descent using $100$ batches, and each batch is formed by $2$ draws from the training simulations. We deliberately choose a small batch size to avoid memory issues and ensure the network's capability to handle large input and output images ($144^3$ and $128^3$ respectively). Selecting a small batch size induces more instability during training; we mitigate this issue by using the clip normalization operation defined in \texttt{TensorFlow} during the backpropagation step.

The hyper-parameter $\beta$ in the Balanced Cross-Entropy \Eq{eq_BaCE} is determined by computing the ratio of negative samples to the total number of samples in the training data. The value of $\beta$ measured in different training simulations lies in the interval $\left[0.575, 0.5892\right]$. There exists a slight predominance of voxels/particles that do not collapse into DM haloes with mass $M_\textrm{FoF} \gtrapprox 10^{11}\,h^{-1}{\rm M_\odot}$ at $z=0$ considering the \cite{2020A&A...641A...6P} cosmology. We fix the hyper-parameter $\beta$ in \Eq{eq_BaCE} to the mean value $\beta = 0.5815$.

Regarding the hyper-parameters in the Weinberger loss \Eq{eq_Weinberger}, we adopt the values presented in \cite{2017arXiv170802551D}, as we have observed that varying these parameters does not significantly affect our final results. The specific hyper-parameter values are the following: $c_{\text{Pull}} = 1$, $\delta_{\text{Pull}}=0.5$, $c_{\text{Push}}=1$, $\delta_{\text{Push}}=1.5$, and $c_{\text{Reg}}=0.001$. We have conducted a hyper-parameter optimization for the clustering algorithm described in Appendix~\ref{sec_A2} and found the following values: $N_{\rm{dens}} = 20$, $N_{\rm{ngb}} = 15$ and $p_{\rm{thresh}} = 4.2$ (see Table~\ref{tab_confusion_matrix}).

Our semantic and instance models are designed to predict regions comprising $128^3$ particles due to technical limitations regarding GPU memory. 
To overcome this limitation and enable the prediction of larger simulation volumes, we have developed an algorithm that seamlessly integrates sub-volume crops. For our semantic model, we serially concatenate sub-volume predictions to cover the full simulation box. For our instance network, we propose the method described in Appendix~\ref{sec_A3}. In summary, this method works as follows: we generate two overlapping lattices. Both lattices cover the entire simulation box, but the second one is shifted with respect to the first one (its sub-volume centres lay in the nodes of the first one). The overlapping regions between the lattices are employed to determine whether instances from different crops should merge or not. We have verified that this procedure is robust by checking that the final predictions are not sensitive to the particular lattice choice.

We train our semantic and instance networks separately. The semantic predictions are not employed at any stage during the training process of the instance model. To compute the instance loss, \Eq{eq_Weinberger} is evaluated using the true instance maps and the pseudo-space positions. The semantic predictions are only employed once both models have been trained. We use the semantic predictions to mask out pseudo-space particles not belonging to haloes. Then, the clustering algorithm described in Appendix~\ref{sec_A2} is applied to identify clusters of particles in the pseudo-space (which yields the final proto-halo regions).

\begin{table}
\caption{Hyper-parameters employed in our instance segmentation pipeline.}
\label{tab:my-table}
\begin{tabular}{ccccccccc}
\hline\hline
$\delta_{\text{Pull}}$ & $\delta_{\text{Push}}$ & $c_{\text{Pull}}$ & $c_{\text{Push}}$ & $c_{\text{Reg}}$ & $N_{\rm{dens}}$ & $N_{\rm{ngb}}$ & $p_{\rm{thresh}}$ \\ \hline
0.5                    & 1.5                    & 1                  & 1                  & 0.001             & 20              & 15             & 4.2               \\ \hline
\end{tabular}
\end{table}

\section{Model Evaluation}\label{sec_Results}

In this section, we test the performance of our models for semantic segmentation (\Sub{subsec_31Semantic}) and instance segmentation (\Sub{subsec_32Instance}). We use the two simulations reserved for validation to generate the results presented in this section.

\subsection{Semantic Results}\label{subsec_31Semantic}

\begin{figure*}
\begin{center}
\includegraphics[width=1.\textwidth]{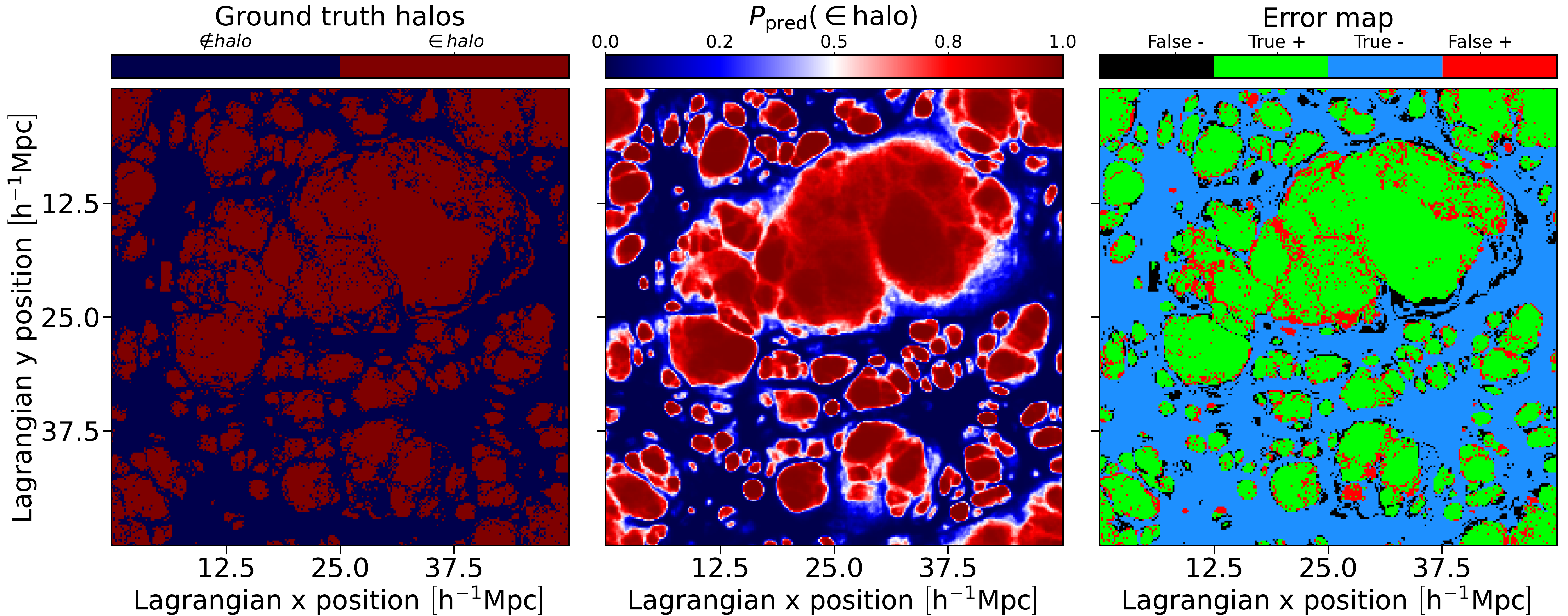}\vspace*{0.cm}
\caption{Slice through the predictions of our semantic segmentation network applied to a validation simulation. \textbf{Left panel}: Ground truth representation showing in red the voxels/particles belonging to a DM halo at $z=0$ and in blue those particles that do not belong to a DM halo. \textbf{Central panel}: Probabilistic predictions of the semantic network with colour-coded probabilities for halo membership. \textbf{Right panel}: Pixel-level error map indicating true positive (green), true negative (blue), false negative (black), and false positive (red) regions resulting after applying a semantic threshold of $0.589$ to our predicted map. The network effectively captures complex halo boundaries and exhibits high validation accuracy ($\textrm{acc}=0.86$) and $\textrm{F}_1$-score ($\textrm{F}_1=0.83$).}
\label{fig_4}
\end{center}
\end{figure*}

In \Fig{fig_4}, we compare the predictions of the semantic segmentation network with the halo segmentation found in the validation simulation. The leftmost panel illustrates a slice of the ground truth. Voxels/particles of the initial conditions belonging to a DM halo at $z=0$ are shown in red; blue voxels represent particles not belonging to a DM halo at $z=0$.

The central panel of \Fig{fig_4} displays the probabilistic predictions from our semantic model for the same slice. The colour map indicates the probability assigned to each pixel for belonging or not to a DM halo. Voxels with a white colour have a $50\%$ predicted probability of belonging to a halo. The neural network tends to smooth out features, assigning uncertain probabilities to regions near halo borders, while consistently assigning high probabilities to inner regions and low probabilities to external regions. In the ground truth it is possible to observe that some interior particles within proto-haloes are predicted to belong to the background. We refer to these as "missing voxels". One of the consequences of the smoothing effect of our network is to ignore these missing voxels, predicting a homogeneous probability of collapse in the interior regions of proto-haloes. The missing voxels in the Lagrangian structure seem to be a feature very sensitive to the initial conditions impossible to capture accurately at a voxel level.  This is supported by the fact that the missing voxels also change significantly in the baseline simulations (see \Fig{fig_1}).

The rightmost panel of \Fig{fig_4} shows the pixel-level error map for the same slice. We select a semantic threshold value equal to $0.589$ to generate these results. We choose this value for the semantic threshold so that the total predicted number of particles that belong to a halo matches the number of collapsed voxels in the validation simulations. In Appendix~\ref{sec_A25} we further analyze the sensitivity of our semantic results to the value chosen for the semantic threshold. We use different colours to represent the corresponding classes of the confusion matrix: Green corresponds to true positive (TP) cases, blue to true negatives (TN), black to false negatives (FN), and red to false positives (FP). 

Some regions are particularly challenging to predict for the network, likely due to their sensitivity to changes in the initial conditions. For example, in the rightmost panel of \Fig{fig_4}, it is easy to appreciate many FN regions that appear as black string-like structures surrounding TP collapsed regions. These FN cases likely correspond to particles infalling into the halo at $z=0$, identified as part of the FoF group despite not having completed the first pericentric passage. Capturing this behaviour might be particularly challenging for the network since the exact shape of these ``first-infall'' regions is more sensitive to small changes in the initial condition and can also be influenced by distant regions of the proto-haloes that do not completely fit within the field-of-view of our network (which can occur for very massive proto-halos). Also, we can appreciate FP regions that appear between the FN string-like regions and the TPs corresponding to the central Lagrangian regions of haloes. Additionally, the boundaries of the largest haloes may be especially difficult to predict for the network, since they only fit partially into the field of view.

The results presented in \Fig{fig_4} suggest, upon visual inspection, that our model accurately captures many of the complex dynamics that determine halo collapse. To rigorously assess the performance of our model we need to quantify the results obtained from our semantic network and compare them with the differences between the baseline simulations, as discussed in \Sec{sec_Methodology}.

In Table~\ref{tab_confusion_matrix} we present the values of some relevant metrics that we can employ to evaluate the performance of our semantic network (we have considered the semantic threshold of $0.589$). In particular, we study the behaviour of five different metrics: True Positive Rate $\mathrm{TPR} = \mathrm{TP}/(\mathrm{TP} + \mathrm{FN})$, True Negative Rate $\mathrm{TNR} = \mathrm{TN}/(\mathrm{TN} + \mathrm{FP})$, Positive Predictive Value $\mathrm{PPV} = \mathrm{TP}/(\mathrm{TP} + \mathrm{FP})$, Accuracy $\mathrm{ACC}$ and the $\textrm{F}_1$-score (which is a more representative score than the accuracy when considering unbalanced datasets), see \Eq{eq_metrics}:

\begin{equation}\label{eq_metrics}
\textrm{ACC} = \frac{\textrm{TP} + \textrm{TN}}{\textrm{TP} + \textrm{TN} + \textrm{FN} + \textrm{FP}} \quad ;\quad  \textrm{F}_1 = \frac{2\textrm{TP}}{2\textrm{TP} + \textrm{FP} + \textrm{FN}}
\end{equation}

Table~\ref{tab_confusion_matrix} also contains the scores measured using the baseline simulations. Our model returns values for all the metrics very close to the optimal target from the baseline simulations. This demonstrates the reliability of our model in predicting the well-specified aspects of halo collapse. See Appendix~\ref{sec_A25} for a more detailed discussion about the performance of our semantic model and the relation between the selected semantic threshold with the results contained in Table~\ref{tab_confusion_matrix}.

In addition to the optimal case, we compare our semantic model with the explicit implementation of the excursion set theory from \textsc{ExSHalos}~\citep{2019arXiv190606630V}. The \textsc{ExSHalos} code grows spheres around the density peaks in the Lagrangian density field until the average density inside crosses a specified barrier for the first time. The barrier shape is motivated by the ellipsoidal collapse~\citep{2001MNRAS.323....1S, 2011MNRAS.418.2403D} with three free parameters that were fitted to reproduce the mean mass function of our simulations. In Table~\ref{tab_confusion_matrix} we include the semantic metrics measured with the \textsc{ExSHalos} results. While \textsc{ExSHalos} can describe halo formation to some degree, there exist some aspects that go beyond the spherical excursion set paradigm which are better captured by our semantic model. A more detailed analysis of the results obtained with \textsc{ExSHalos} is presented in Appendix~\ref{sec_A4}.

\begin{table}
\begin{center}
\caption{Performance metrics of our semantic segmentation model, along with the \textsc{ExSHalos} results, compared against the optimal target accuracy estimated from the baseline simulations. The table presents True Positive Rate (TPR), True Negative Rate (TNR), Positive Predictive Value (PPV), and Negative Predictive Value (NPV).}
\label{tab_confusion_matrix}
\begin{tabular}{lccccc}
\hline\hline
Type  & TPR   & TNR   & PPV   & ACC   & $\textrm{F}_1$ \\ \hline
\textsc{ExSHalos} & 0.518 & 0.845 & 0.707 & 0.708 & 0.598\\ 
Pred. & 0.838 & 0.883 & 0.838 & 0.864 & 0.838    \\ 
Optimal & 0.887 & 0.914 & 0.882 & 0.903 & 0.884              \\ \hline
\end{tabular}
\end{center}
\end{table}

\begin{figure}
  \resizebox{\hsize}{!}{\includegraphics{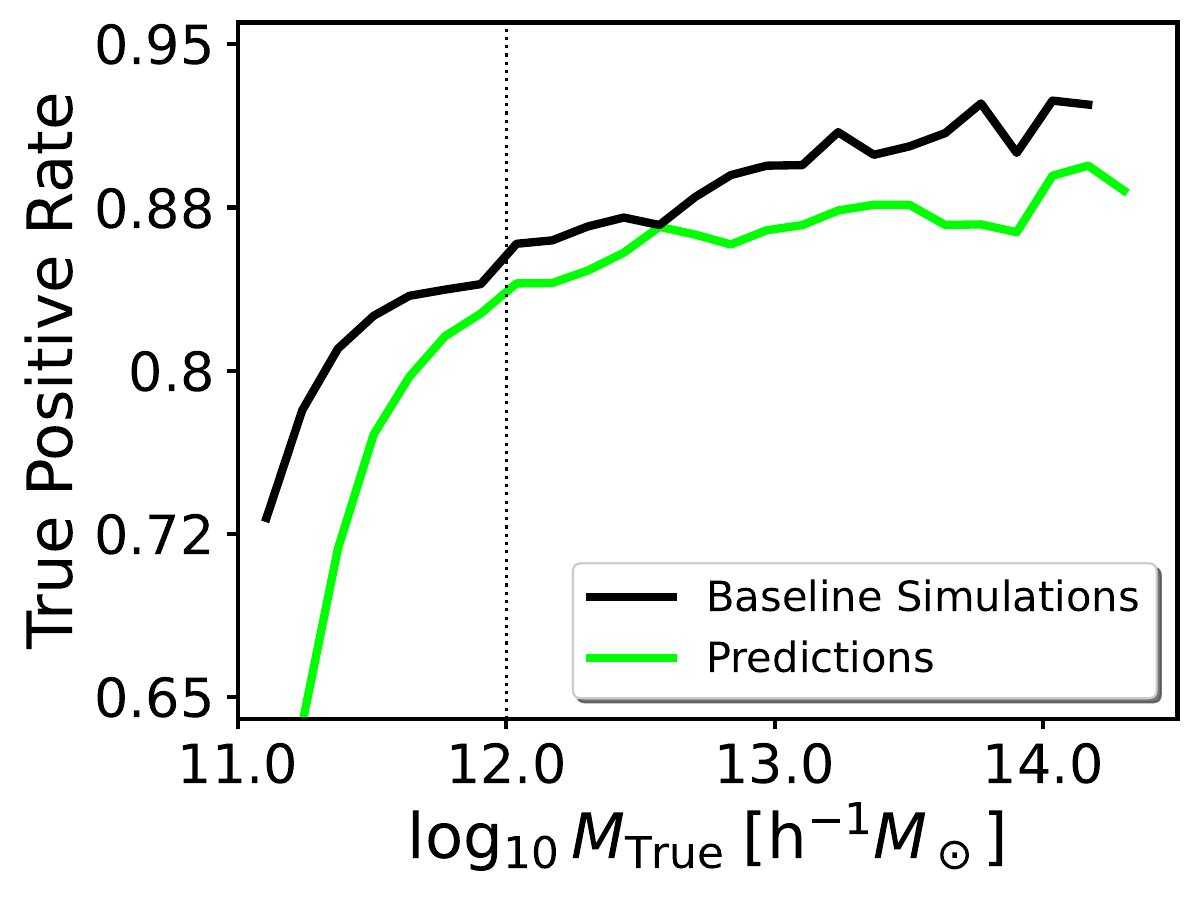}}
  \caption{True Positive Rate expressed as a function of the halo mass associated with the ground truth voxels. We present the results measured from the model predictions (solid bright green line) in comparison to the optimal target accuracy from the baseline simulations (solid black line). The vertical dotted line at $10^{12}\,h^{-1}{\rm M_\odot}$ marks the point where model predictions start to differ from the baseline results.}
  \label{fig_5}
\end{figure}

In \Fig{fig_5} we compare the values of the predicted TPR as a function of ground truth halo mass ($\textrm{TPR}_\textrm{Pred}$, solid green line), with the TPR values measured from the baseline simulations ($\textrm{TPR}_\textrm{base}$, solid black line). It is possible to perform this comparison for the TPR because, in the ground truth data, we retain information about the mass of the FoF-haloes associated with each DM particle. Therefore, we can compute the fraction of TP cases in different ground-truth-mass-bins by selecting the voxels according to the mass associated with them in the ground truth.

In \Fig{fig_5}, the values for $\textrm{TPR}_\textrm{base}$ increase with halo mass, indicating that particles that end up in lower-mass haloes are more sensitive to small-scale changes in the initial conditions, consequently, harder to predict accurately. Our network's predictions follow a similar trend, albeit with some discrepancies. The model seems to under-predict the number of particles that end up in haloes with masses lower than $M_\textrm{True} \lessapprox 10^{12}\,h^{-1}{\rm M_\odot}$ (dotted vertical black line in \Fig{fig_5}). This indicates that our model tends to under-predict the number of pixels that are identified as TPs in the lower mass end. For haloes whose mass is greater than $10^{12}\,h^{-1}{\rm M_\odot}$, our model returns accurate predictions to a good degree over a broad range, extending more than two orders of magnitude in halo mass.

In this subsection, we have demonstrated that our semantic model extracts most of the predictable aspects of halo formation by comparing our results with the baseline simulations (which only differ in unresolved aspects of the initial conditions). We now employ the predictions of our semantic network to generate the final results using our instance segmentation model.

\subsection{Instance Results}\label{subsec_32Instance}

\begin{figure}
  \resizebox{\hsize}{!}{\includegraphics{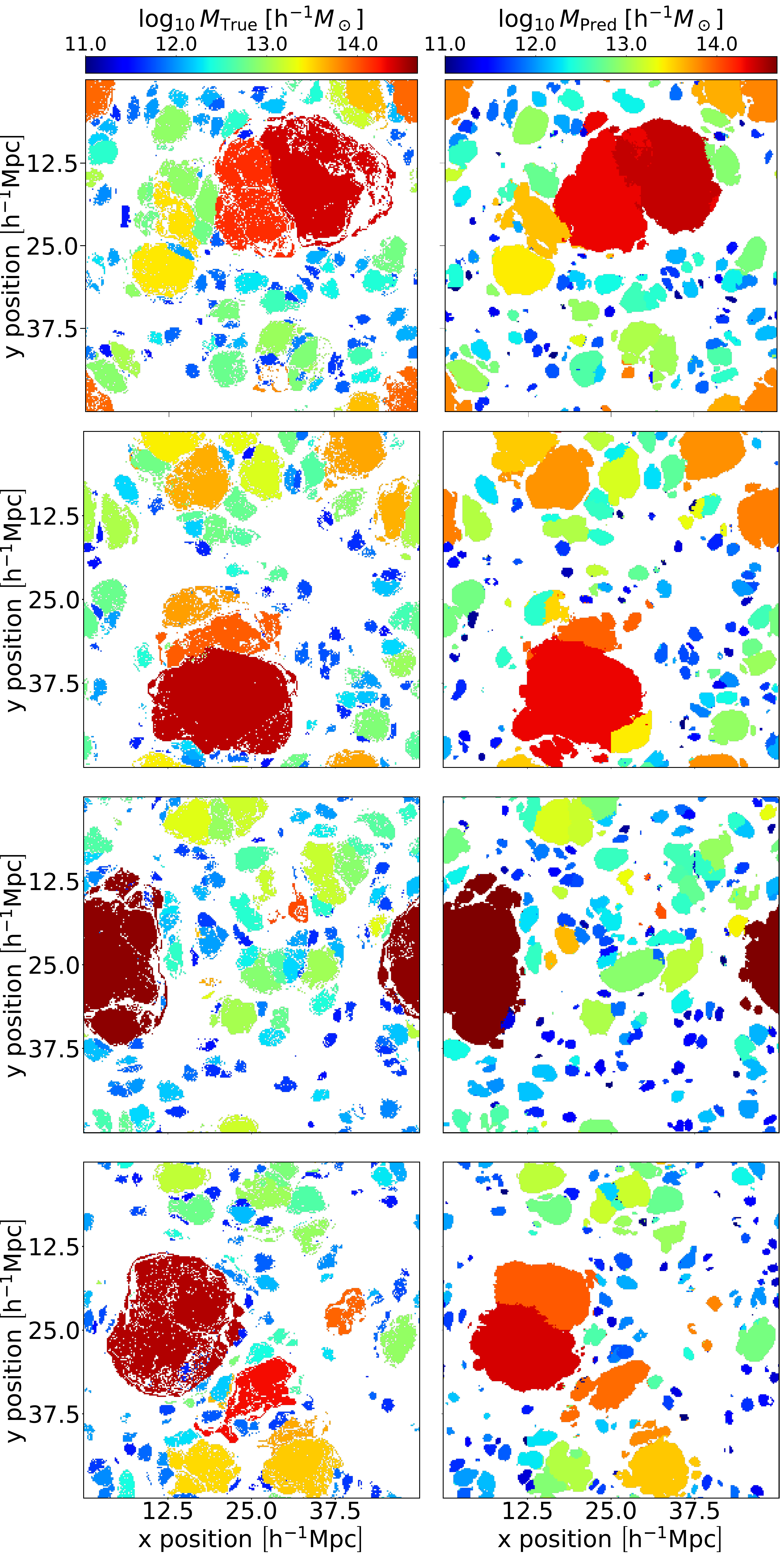}}
  \caption{Examples of the instance segmentation results obtained with our model. \textbf{Left column}: ground truth masses obtained using N-body simulations.\textbf{ Right column}: predicted masses obtained using our instance segmentation pipeline. The model can predict the Lagrangian patches of haloes, although some small differences -- e.g. regarding the connectivity of haloes -- exist.}
  \label{fig_6}
\end{figure}

We provide some examples of our instance predictions in \Fig{fig_6}. The left column displays the ground truth masses of halo Lagrangian regions extracted from the simulation results (analogous to \Fig{fig_1}); the right column shows the predictions obtained from our segmentation pipeline. The way in which we compute halo masses from the instance predictions is by counting the number of particles/voxels that have been assigned to the same label and multiplying that by the particle mass of our simulations, $m_{\rm DM}=6.35\cdot10^{8}\,h^{-1}{\rm M_\odot}$.

The shapes of the halo contours are well-captured thanks in part to the semantic predictions. The instance segmentation pipeline successfully distinguishes the different haloes that have formed, and in most cases, correctly separates neighbouring haloes. This is not a trivial task since the size of halo Lagrangian regions varies across several orders of magnitude. Therefore, the instance segmentation pipeline must correctly separate wildly different particle groupings in the pseudo-space. \Fig{fig_6} shows that our instance segmentation pipeline correctly identifies different Lagrangian halo regions for the majority of cases. However, we note that differences arise on the one hand for very small haloes that are close to the resolution limit and on the other hand for very large haloes that are larger than the field of view of the network.

\begin{figure}
  \resizebox{\hsize}{!}{\includegraphics{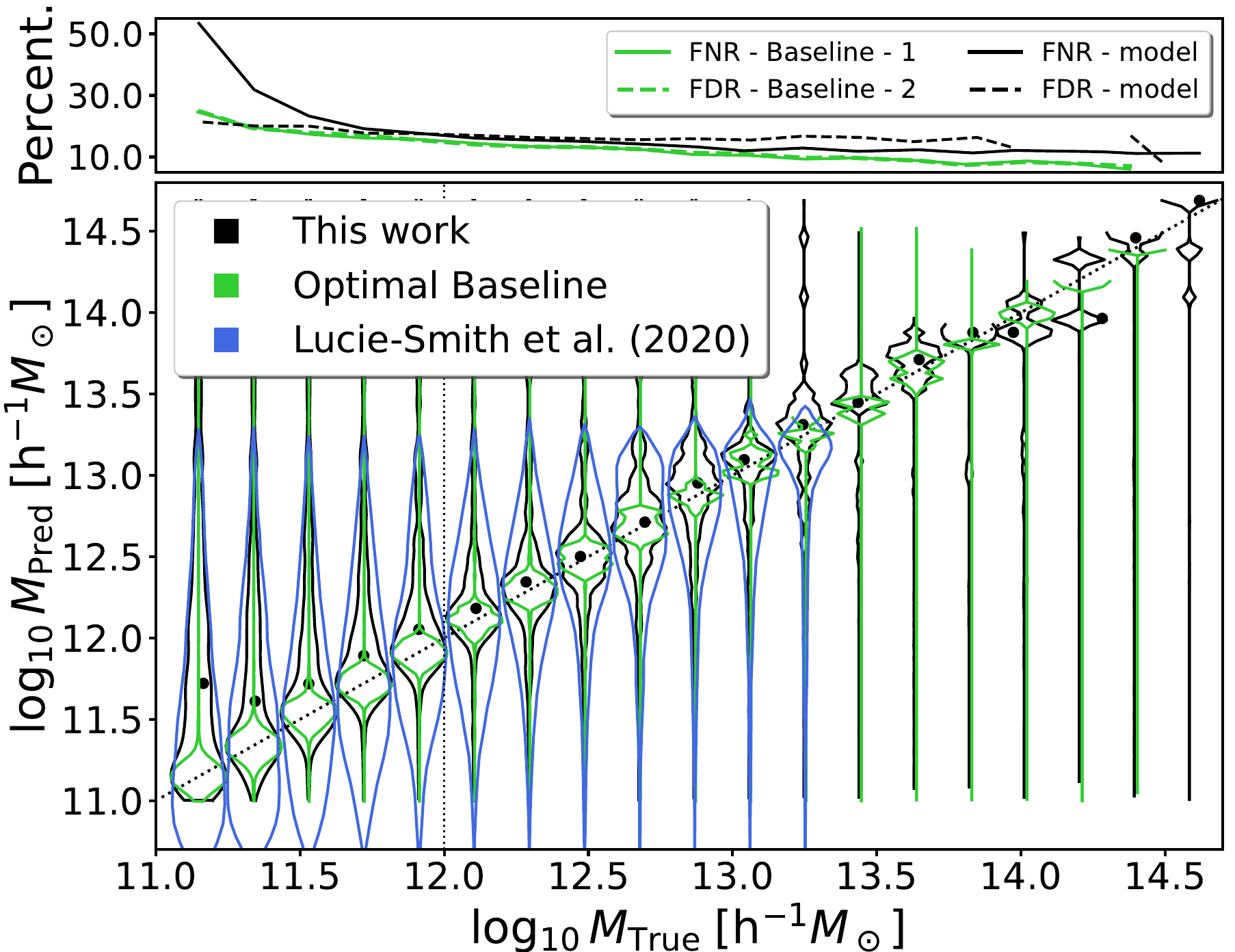}}
  \caption{``Violin plot'', visualizing the distribution of predicted halo masses (at a voxel level) for different ground-truth mass bins. The black violin plots show the results obtained with our instance segmentation model. Green violin plots show the agreement between the two baseline simulations -- representing an optimal target accuracy. The blue violin plots in the main panel show the results presented in \citep{2020arXiv201110577L}. The solid black line in the top panel shows the false negative rate, FNR, as a function of the ground truth halo mass. The dashed black line represents the fraction of predicted collapsed pixels that are not actually collapsed as a function of predicted halo mass (false discovery rate, $\textrm{FDR}$). The green lines on the top panel correspond to the analogous results obtained from the baseline simulations. The model predicts haloes accurately object-by-object for masses $M \gtrsim 10^{12} M_\odot /h$.}
  \label{fig_7}
\end{figure}

In \Fig{fig_7}, we present a comparison between the ground truth halo masses and the predicted masses associated with the particles/voxels in our validation set. To generate these results we apply the following procedure: We select all the ground truth voxels/particles that end up in FoF-haloes and study the predictions associated with them. We can associate a predicted mass for all the voxels that belong to a DM halo. In these cases, we can compare the predicted mass values ($M_{\mathrm{Pred}}$) with the ground truth masses ($M_{\mathrm{True}}$) at a voxel level. This comparison is shown in the main panel of \Fig{fig_7} as black violin plots (``violins'' henceforth). The mass range covered by the black violins goes from $M_{\mathrm{True}}=10^{11}\,h^{-1}\rm M_\odot$, corresponding to the minimum mass of haloes ($155$ particles), to $M_{\mathrm{True}}\approx10^{14.7}\,h^{-1}\rm M_\odot$, which is the mass of the most massive halo identified in the validation simulations. The number of high-mass haloes is smaller than small-mass ones and therefore the higher-mass end of the violin plot exhibits more noise. We can appreciate that the median predictions (black dots) correctly reproduce the expected behaviour (ground truth) for several orders of magnitude.

The voxels identified as part of a halo in the ground truth, but not in the predicted map, are false negative (FN) cases. For these occurrences, we can study the dependence of the False Negative Rate (FNR) as a function of the ground truth halo mass (solid black line on the top panel of \Fig{fig_7}; analogous to \ref{fig_5}). We can also study the reciprocal case in which a voxel is predicted to be part of a halo (hence, it has an associated $M_{\mathrm{Pred}}$) but the ground truth voxel is not collapsed. These cases correspond to False positives (FP) but to make a comparison as a function of mass we can only express it in terms of the predicted mass. Therefore, we show as a dashed black line in the top panel of \Fig{fig_7} the false discovery rate,
\begin{equation}\label{eq_FDR}
\textrm{FDR} = \frac{\left[\textrm{FP} |M_\mathrm{Pred}\right]}{\left[\textrm{TP} |M_\mathrm{Pred}\right] + \left[\textrm{FP} |M_\mathrm{Pred}\right]} \mathrm{\quad .}
\end{equation}

We compare our results with those obtained from the baseline simulations. In the main panel of \Fig{fig_7} we present the corresponding violin plots from the baseline simulations with green lines. The range that the green violins span is smaller than the black violins since the most massive halo identified in the baseline simulations has a mass of $M_{\mathrm{True}}\approx 10^{14.4}\,h^{-1}\rm M_\odot$. In the top panel, the solid and dashed green lines represent the FPR and $\textrm{FDR}$ respectively. As expected, the FPR and $\textrm{FDR}$ coincide in the case of the baseline simulations. The top panel results demonstrate that our predictions are comparable to those of the baseline simulations (as pointed out in \Fig{fig_5}) over most of the considered mass range. However, they get progressively worse for masses below $M_\textrm{True} \lessapprox 10^{12}\,h^{-1}{\rm M_\odot}$ (vertical dotted black line), deviating from the baseline trend. This indicates that our model struggles to capture the correct behaviour of lower-mass haloes but it produces accurate predictions for higher-mass ones. When comparing the violin plot distributions of our model with the baseline simulations we appreciate that we obtain similar (but slightly broader)  contours. Being able to achieve a similar scatter as in the baseline simulations indicates that our model can capture the well-resolved aspects of halo formation. We want to emphasize that precise predictions for halo masses are not directly enforced through the training loss, but are a side product, consequence of precisely reproducing halo Lagrangian patches. The scatter broadens for smaller halo mass and the network loses accuracy in these cases, sometimes associating smaller haloes close to a big Lagrangian patch to its closest more massive neighbour.

In the main panel of \Fig{fig_7}, we include the violin plot lines presented in~\cite{2020arXiv201110577L} (blue violin lines). In this study, a neural network was trained to minimize the difference between predicted and true halo-masses at the particle level using as inputs the initial density field or the potential. The focus of \cite{2020arXiv201110577L} is to examine how different features of the initial conditions influence mass predictions within a framework that mirrors analytical models. 

The comparison between our methodology and \cite{2020arXiv201110577L} in \Fig{fig_7} highlights the differing outcomes that arise from the unique objectives and constraints each model employs. While both models ultimately predict halo masses, we suggest that our approach benefits from the rigid operator that groups particles together and assigns them the same halo mass. Therefore, analytical approaches towards predicting the formation of structures may benefit from knowing about the fate of neighbouring particles. Since in excursion set formalisms, this is only possible to a limited degree, this increases the motivation for considering alternative approaches, like the one proposed by~\cite{2023MNRAS.523L...4M}.

\begin{figure}
  \resizebox{\hsize}{!}{\includegraphics{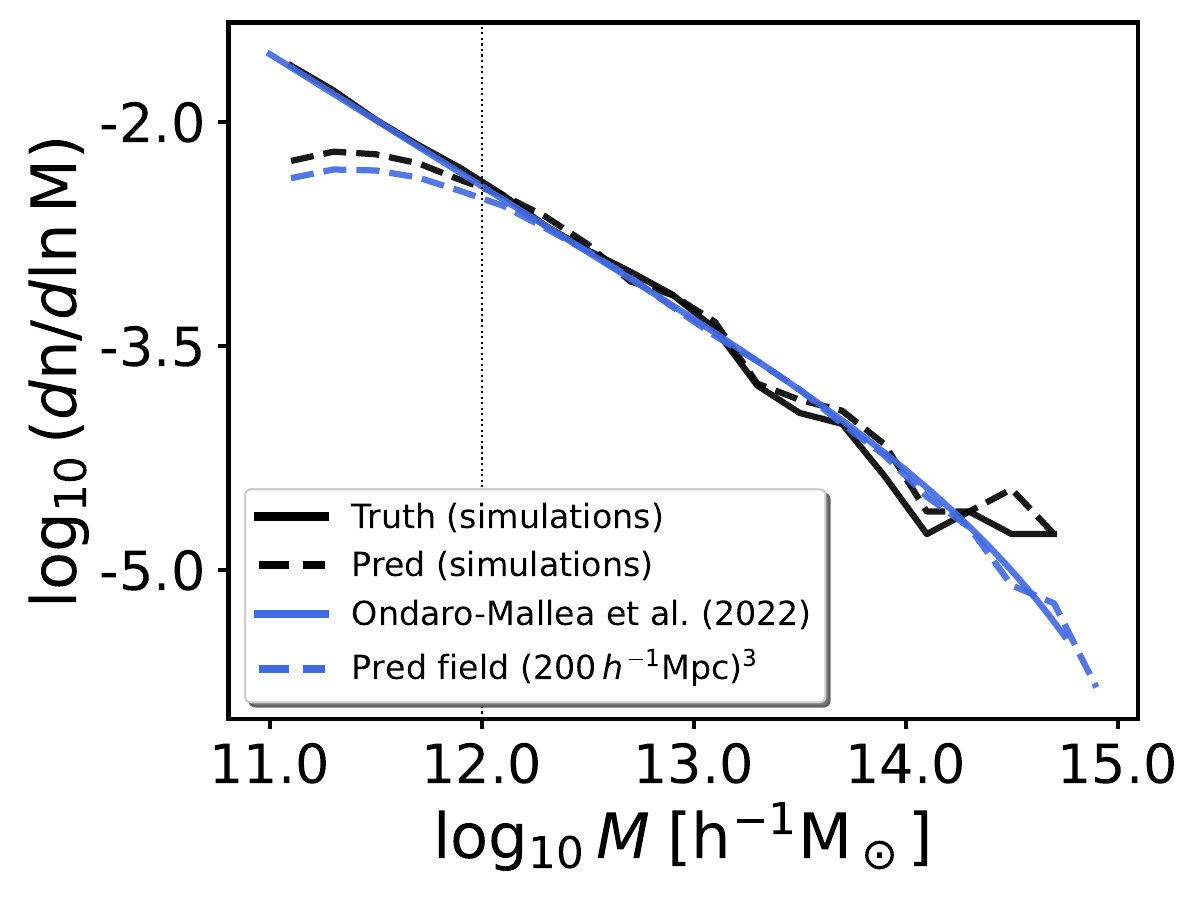}}
  \caption{Halo-mass-function (HMF) computed using our N-body simulations reserved for validation (solid black line). The dashed black line represents the predicted HMF using the Lagrangian halo regions obtained with our instance segmentation pipeline. The solid blue line shows the HMF prediction from~\citep{2022MNRAS.509.6077O}. The dashed blue line corresponds to the HMF obtained after evaluating our model in a simulation with $1024^3$ particles and $V_{\textrm{box}} = (200\,h^{-1}\mathrm{Mpc})^3$.}
  \label{fig_HMFs}
\end{figure}

In Appendix~\ref{sec_A4}, we include a comparison of our instance model with the predictions of \textsc{ExSHalos}~\citep{2019arXiv190606630V}. In \Fig{fig_ExSHalos_map}, we show a map-level comparison between the Lagrangian shapes of friends-of-friends proto-haloes and \textsc{ExSHalos} predictions. The shapes of proto-haloes predicted by the \textsc{ExSHalos} implementation are limited to sphere-like volumes, which affects its flexibility and, consequently, its accuracy (see Table~\ref{tab_confusion_matrix}). While \textsc{ExSHalos} correctly replicates the halo mass function of friends-of-friends haloes, it  struggles to reproduce particle-level mass predictions, as shown in the violin plot in \Fig{fig_ExSHalos_violin}.

In \Fig{fig_HMFs} we present the halo-mass-function (HMF) computed using the validation simulations (solid black line). The dashed black line shows the predicted HMF computed using the results of our instance segmentation pipeline. We can appreciate that our predictions reproduce the N-body results over a range that spans more than two orders of magnitude. Our results improve upon the prediction mass range for the HMF of previous similar approaches~\citep{2019MNRAS.482.2861B, 2020MNRAS.496.5116B}. This is despite the fact that ~\cite{2020MNRAS.496.5116B} select their hyper-parameters to reproduce the HMF; while in~\cite{2019MNRAS.482.2861B} they reproduce the HMF corresponding to \texttt{Peak Patch} haloes~\citep{2019MNRAS.483.2236S}, instead of the HMF associated with FoF haloes. In \Fig{fig_HMFs} we also include a solid blue line representing the theoretical HMF predictions using the model by~\cite{2022MNRAS.509.6077O}. We compare this result with the HMF associated with the haloes predicted by our model using the density and potential fields of a realization with $1024^3$ particles and a volume of $V_{\textrm{box}} = (200\,h^{-1}\mathrm{Mpc})^3$. Both lines show a good agreement in the $10^{12} - 10^{15} \,h^{-1}{\rm M_\odot}$ range.

We conclude that our semantic plus instance segmentation pipeline correctly reproduces the Lagrangian halo shapes of FoF-haloes spanning a mass range between $10^{12}\,h^{-1}{\rm M_\odot}$ and $10^{14.7}\,h^{-1}{\rm M_\odot}$. We have tested the accuracy of our results employing different metrics (presented in several tables and figures). Inferred quantities from our predicted Lagrangian halo regions, such as the predicted halo masses, correctly reproduce the trends computed using N-body simulations and improve upon the results presented in previous studies.

\section{Experiments}\label{sec_Experiments}

In this section, we test how our network reacts to systematic modifications to the input density field and potential and how well it generalizes to scenarios that lie beyond the trained domain. Therefore, we analyze the response to large-scale density perturbations, to large-scale tidal fields and to changes in the variance of the density field.

\subsection{Response to large scale densities} \label{subsec_41_exp_dens}

We study the response of the haloes to a large-scale over-density such as typically considered in separate universe simulations \citep{wagner_2015, lazeyras_2016, li_2014}. We add a constant $\delta_\epsilon$ to the input density field $\delta(\vec{q})$ so that the new density field $\delta_*(\vec{q})$ is given by
\begin{align}
    \delta_*(\vec{q}) &= \delta(\vec{q})  + \delta_\epsilon,
\end{align}
and to maintain consistency with Poisson's equation, see \Eq{eq_potential}, we add a quadratic term to the potential:
\begin{align}
    \phi_*(\vec{q}) &= \phi(\vec{q}) + \frac{\delta_\epsilon}{6} (\vec{q} - \vec{q}_0)^{2} \label{eqn:pot_from_density_pert}
\end{align}
where $\vec{q}_0$ is an arbitrary (and irrelevant) reference point \citep{stucker_2021}, which we choose to be in the centre of our considered domain. Note that we break the periodic boundary conditions here, so it is difficult to do this operation for the whole box, but instead we consider it only for a smaller region to avoid boundary effects.

\begin{figure}
\begin{center}
  \resizebox{.7\hsize}{!}{\includegraphics{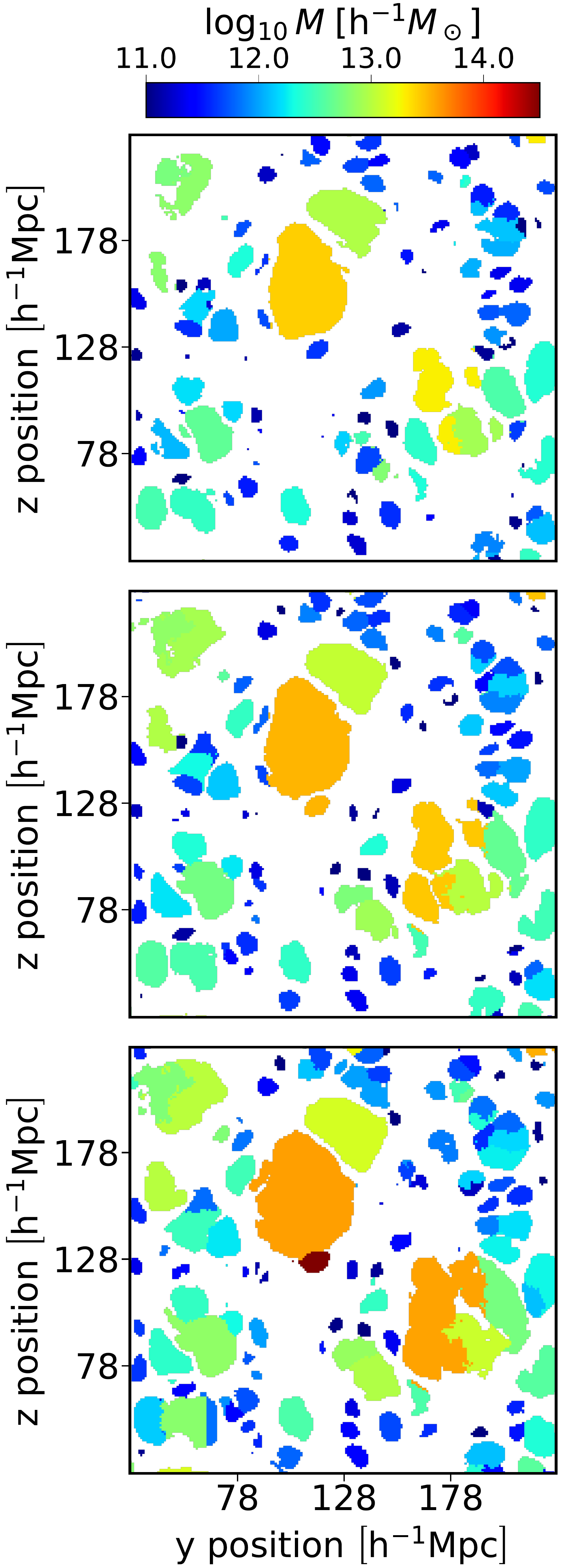}}
  \caption{Response of proto-haloes to large-scale over-densities. The three panels show over-densities of $\delta_\epsilon = -0.5$, $0$ and $0.5$ respectively. A larger large-scale density tends to increase the Lagrangian volume of haloes and leads to additional mergers in some cases.}
  \label{fig_Figure_experiment_density}
\end{center}
\end{figure}

We show how haloes respond to this modification in \Fig{fig_Figure_experiment_density}. The middle panel shows the predicted masses associated with the particles/voxels (in a similar way to \Fig{fig_6}) for the reference field, $\delta_\epsilon = 0$. The upper and lower panels show the results of including a constant term to the initial over-density field of $\delta_\epsilon=-0.5$ and $\delta_\epsilon=0.5$, respectively.

Increases in the background density lead to more mass collapsing onto haloes, thus generally increasing the Lagrangian volume of haloes. Furthermore, it leads in many cases to previously individual haloes merging into one bigger structure. This is qualitatively consistent with what is observed in separate universe simulations \cite[e.g.][]{2015JCAP...10..059D, 2015MNRAS.448L..11W, 2019MNRAS.488.2079B, 2019PhRvD.100l3528J, 2022PhRvD.106h3504T, 2022JCAP...02..001A}. 

\begin{figure}
  \resizebox{\hsize}{!}{\includegraphics{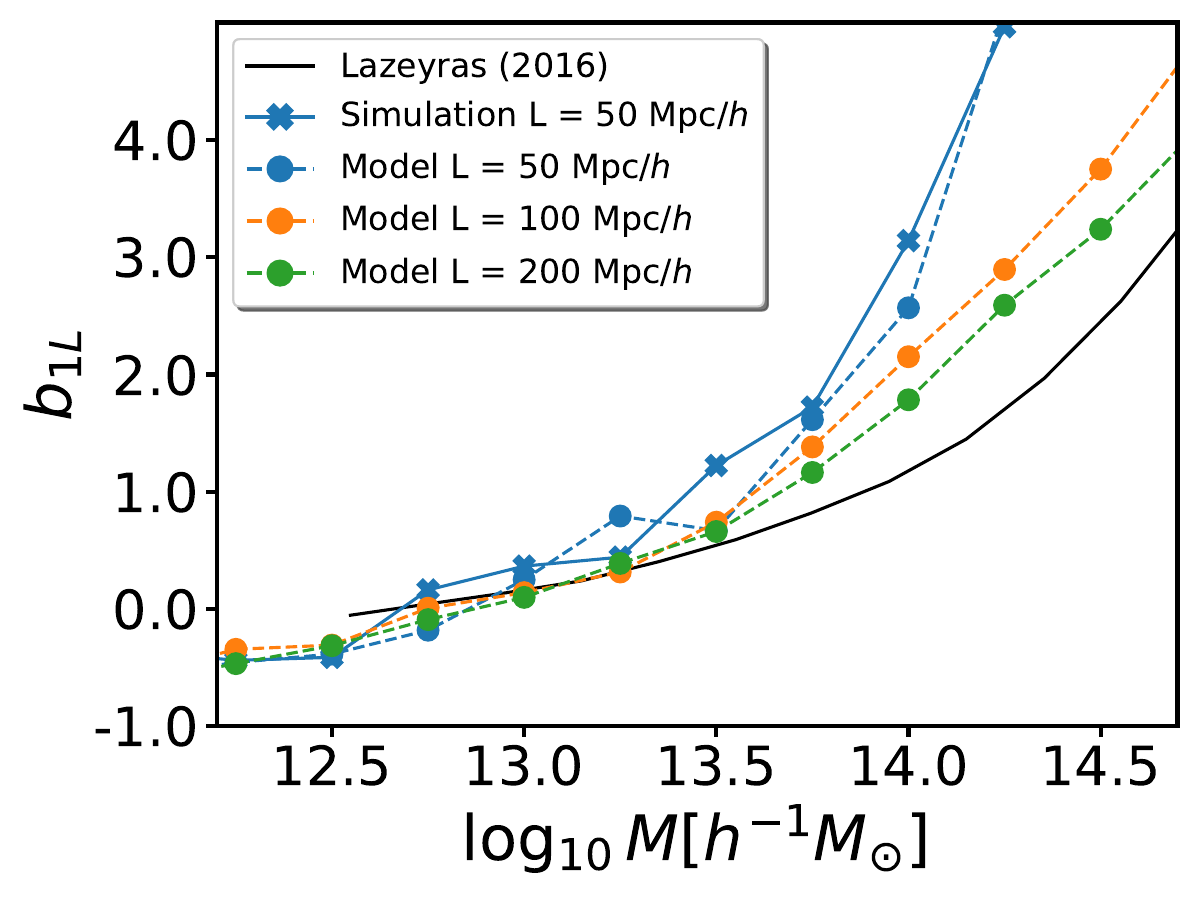}}
  \caption{Linear Lagrangian bias parameter $b_{1L}$ for the haloes, measured for different boxsizes $L$ and comparing simulation and model. The model agrees well with the simulation at the $L = 50 h^{-1}\textrm{Mpc}$ scale, but both are inconsistent with the true large-scale bias relation from \citep{lazeyras_2016} due to effects from the limited size of the simulation volume. Evaluation on larger boxes moves the prediction closer to the known relation, but some deviation is maintained.}
  \label{fig:b1_measurement}
\end{figure}

To evaluate quantitatively whether the model has learned the correct response to large-scale density perturbations, we test whether it recovers the same halo bias that has been measured in previous studies \citep[][for a review]{2018PhR...733....1D}. In separate universe experiments, the linear bias parameter can be inferred as the derivative of the halo mass function with respect to the large-scale density:
\begin{align}
    b_{1L}(M) =  \frac{1}{n_h(M)} \frac{\partial n_h(M) }{\partial \delta_\epsilon} 
\end{align}
Therefore, \citep{lazeyras_2016} used the halo mass function measured in separate universe simulations with different large-scale densities $\delta_\epsilon$ to measure the bias parameters through a finite differences approach. While our qualitative experiment from Figure \ref{fig:b1_measurement} follows this in spirit, it is difficult to do the same measurement here, since the addition of the quadratic potential term in equation \eqref{eqn:pot_from_density_pert} breaks the periodic boundary conditions and makes it difficult to measure the mass function reliably over a large domain. Therefore, we instead adopt an approach to infer the bias from the unperturbed $\delta_\epsilon = 0$ case. \citep{2013MNRAS.436..449P} shows that the Lagrangian bias parameter can be measured by considering the (smoothed) linear over-density at the Lagrangian location of biased tracers $\delta_i$:
\begin{align}
    b_{1L} &= \frac{1}{N} \sum \frac{\delta_i}{\sigma^2} \label{eqn:momentbias}
\end{align}
where the sum goes over $N$ different tracers (e.g. all haloes in a given mass bin) and where $\sigma^2 = \langle \delta^2 \rangle$  is the variance of the (smoothed) linear density field. Since this measurement should give meaningful results only on reasonably large scales, we smooth the Lagrangian density field with a Gaussian kernel with width $\sigma_r = 6 h^{-1}\mathrm{Mpc}$. We measure the smoothed linear density $\delta_i$ at the Lagrangian centre of mass of each halo patch and then we measure the bias by evaluating equation \eqref{eqn:momentbias} in different mass bins. 

We show the resulting $b_{1L}$ as a function of mass in Figure \ref{fig:b1_measurement}. The blue solid and dashed lines show the bias parameters measured in an $L = 50 h^{-1}\textrm{Mpc}$ box for the simulated versus predicted halo patches respectively. These two seem consistent, showing that the model has correctly learned the bias relation that is captured inside of the training set. However, this ($L = 50 h^{-1}\textrm{Mpc}$) relation is not consistent with the well-measured relation from larger scale simulations, indicated as a black solid line adopted from \citep{lazeyras_2016}. This is because very massive haloes $M \gg 10^{14} h^{-1} M_\odot$ do not form in simulations of such a small volume, but they are important to get the correct bias of smaller mass haloes, since wherever a large halo forms, no smaller halo can form. Our network has never seen such large scales, so it is questionable whether it has any chance of capturing the large-scale bias correctly. However, it might be that what it has learned in the small-scale simulation transfers to larger scales. To test this, we evaluate the network on two larger boxes, $L = 100 h^{-1}\textrm{Mpc}$ and $L = 200 h^{-1}\textrm{Mpc}$, shown as orange and green lines in Figure \ref{fig:b1_measurement}. These cases match the true bias relation better, but still show some significant deviation e.g. at $M \sim 10^{14} h^{-1} M_\odot$. Therefore, we conclude that the network generalizes only moderately well to larger scales and halo masses. Improved performance could possibly be achieved by extending the training set to larger simulations and by increasing the field of view of the network.

\subsection{Response to large scale tidal fields}\label{subsec_42_exp_tidal}

\begin{figure}
\begin{center}
  \resizebox{.7\hsize}{!}{\includegraphics{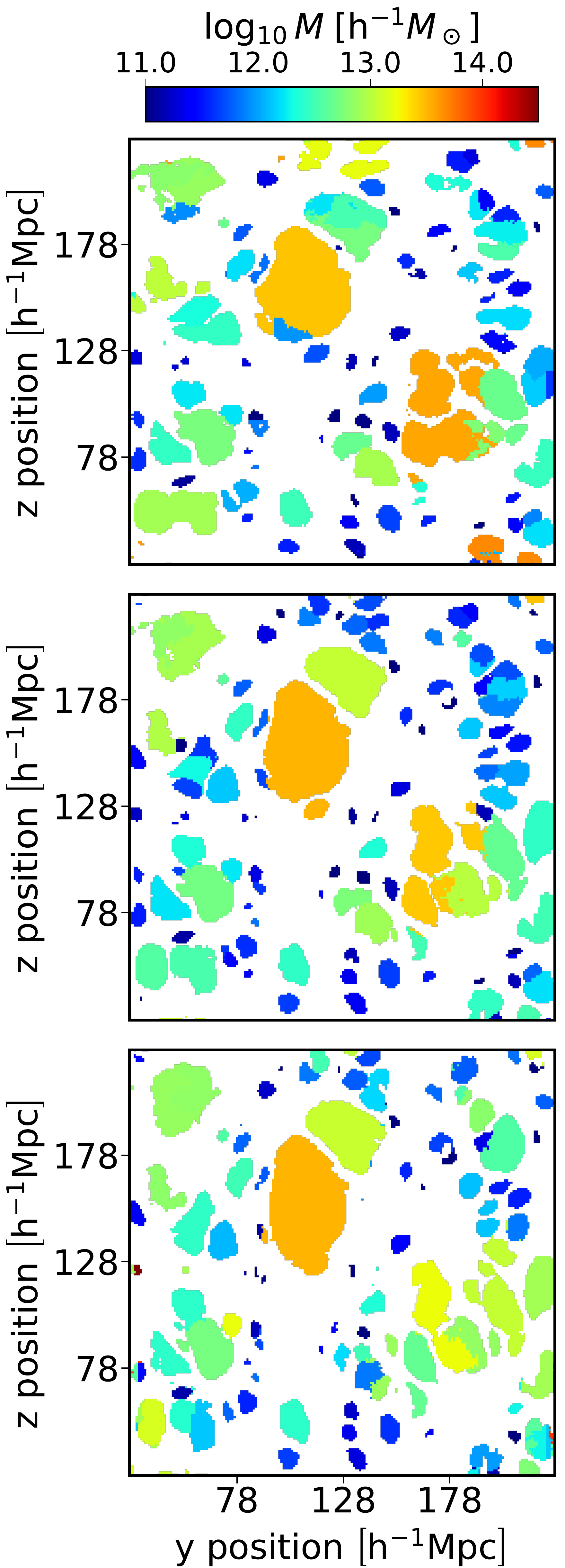}}
  \caption{Response of proto-halo regions towards a large-scale tidal field. The different panels show the cases with $\lambda_z = -0.5$, $0$ and $0.5$ -- corresponding to a stretching tidal field, no tidal field and a compressing tidal field in the vertical direction respectively. A negative (stretching) tidal field delays infall and shrinks the proto-halo patches in the corresponding direction, whereas a positive (compressing) tidal field facilitates infall and extends the proto-halo patches.}
  \label{fig_Figure_experiment_potential}
\end{center}
\end{figure}

In a second experiment, we want to study the response of haloes to purely anisotropic changes of the initial conditions, by adding a large-scale tidal field. We, therefore, aim to emulate a modification similar to the ones considered in anisotropic separate universe simulations \citep{schmidt_2018, 2021MNRAS.503.1473S, masaki_2020, akitsu_2021}. We modify the input potential through the term
\begin{align}
    \phi_*(\vec{q}) &= \phi(\vec{q}) + \frac{1}{2}(\vec{q} - \vec{q}_0)^T T (\vec{q} - \vec{q}_0) \\
     T &= \begin{pmatrix}
        0 & 0 & 0 \\
        0  & - \lambda_z & 0 \\
        0 & 0 & \lambda_z \\
    \end{pmatrix}
\end{align}
Since we are considering a trace-free tidal tensor, we do not need to include any modifications to the initial density field. The results of introducing the tidal field are presented \Fig{fig_Figure_experiment_potential}. In the upper panel in which we have imposed a value of $\lambda_z = -0.5$, the regions of typical proto-haloes are slightly reduced in the $z$-direction and extended in the $y$-direction. Further, in some cases haloes merge additionally in the y-direction while separating in the z-direction. In the bottom panel with $\lambda_z = 0.5$ we observe the opposite behaviour, with proto-halo shapes elongated in the z-direction and reduced in the y-direction. These observations are consistent with the naive expectation: A positive $\lambda_z$ means a contracting tidal field in the z-direction, which facilitates infall in this direction, whereas a negative $\lambda_z$ delays the infall. Therefore, proto-haloes appear extended in the direction where the tidal field has a contracting effect. This should not be confused with the response of the halo shapes in Eulerian space which has the opposite behaviour -- reducing the halo's extent in the direction where the tidal field is contracting \citep{2021MNRAS.503.1473S}. Therefore, a large-scale tidal field effects that \emph{the direction from which more material falls in, is the direction where the final halo is less extended}.

However, by comparing Figures \ref{fig_Figure_experiment_density} and \ref{fig_Figure_experiment_potential}, we note that the effect of modifying the eigenvalues of the tidal tensor (while keeping the trace fixed) is much less significant than modifying its trace $\delta$ by a similar amount. Modifying $\delta$ leads to strong differences in the abundance and the masses of haloes whereas the modifications to the tidal field strongly affect the shapes, but has a much smaller effect on typical masses -- if at all. 

Our investigation into the role of anisotropic features in the initial conditions complements the findings of \cite{2020arXiv201110577L}. They find that anisotropic features of the initial conditions do not significantly enhance halo mass predictions when compared to predictions based on spherical averages. Therefore, they conclude that including anisotropic features would not significantly improve the mass predictions that can be obtained within excursion set frameworks.
This observation is consistent with masses not changing significantly when applying a large-scale tidal field. However, we find that anisotropic features are in general important for the formation of structures since they affect which particles become part of which halo.

Finally, we note that the response of the Lagrangian shape of haloes is particularly interesting in the context of tidal torque theory \citep{white_1984}. To predict the angular momentum of haloes, tidal torque theory requires knowledge of both the tidal tensor and the Lagrangian inertia tensor of haloes. Further, it has been argued that the misalignment of tidal field and Lagrangian inertia tensor is a key factor for predicting galaxy properties \citep{2023arXiv231103632M}. Our experiments show that modifications of the tidal tensor itself also trigger modifications of the Lagrangian shape. Precisely understanding this relation would be relevant to correctly predict halo spins from the initial conditions. Note that such responses are inherently absent in most density-based structure formation models \citep[e.g.][]{1974ApJ...187..425P, 1991ApJ...379..440B, 2002MNRAS.329...61S}, but could possibly be accounted for by recently proposed approaches based on the Lagrangian potential \citep{musso_2021, musso_2023}.

\subsection{Response to changes in the variance of the density field}\label{subsec_43_exp_s8}

\begin{figure}
  \resizebox{\hsize}{!}{\includegraphics{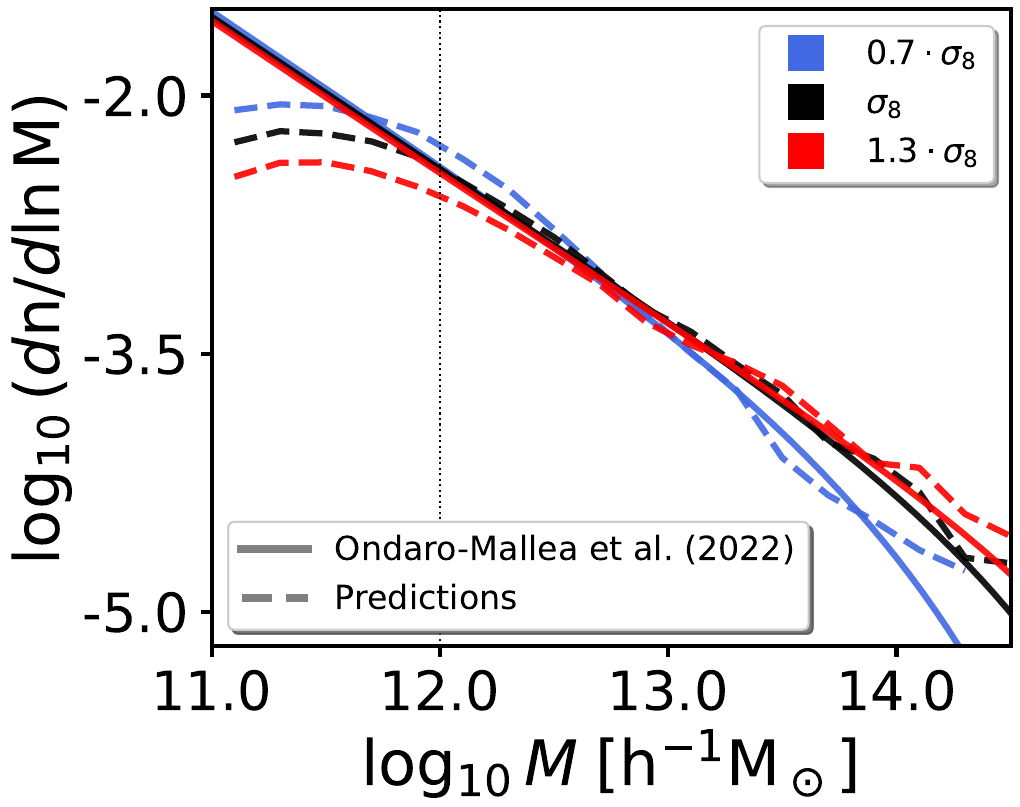}}
  \caption{Comparison of HMF predictions with variations in the cosmological parameter $\sigma_8$. Solid lines represent HMF predictions from~\citep{2022MNRAS.509.6077O}. Dashed lines indicate our model's predictions. Blue and red curves correspond to scenarios with $\sigma_8 = 0.5802$ and $\sigma_8 = 1.077$ respectively. Black lines show the results for $\sigma_8 = 0.8288$ (our reference cosmology).}
  \label{fig_Figure_experiment_s8}
\end{figure}

We now study whether our model can generalize to scenarios different from the training set by investigating how it responds to variations in $\sigma_8$, deviating  $30\%$ from the original \cite{2020A&A...641A...6P} cosmology. We aim to discern if the network, trained on a singular variance setting, has gained enough insight into halo formation to anticipate outcomes considering different values for the variance of the initial density field. These modifications only affect the initial conditions which are fully visible to the network, so it could be possible that the network correctly extrapolates to these scenarios. 

In \Fig{fig_Figure_experiment_s8} we show how the HMF reacts to changes in $\sigma_8$ in comparison to the measured mass functions from~\cite{2022MNRAS.509.6077O} (solid lines) as a benchmark. Our predictions for the HMF (dashed lines) are generated by taking the average results of $10$ different boxes, each one spanning $L = 50 h^{-1}\textrm{Mpc}$, with $\sigma_8$ values set to $0.5802$ (blue lines), $0.8288$ (black lines), and $1.077$ (red lines). The model's predictions reveal a discrepancy with the anticipated HMF behaviour beneath the threshold of $\sim 10^{12.7} h^{-1} M_\odot$ for both $\sigma_8 \approx 0.5802$, and $\sigma_8 \approx 1.077$. This discrepancy is attributed to the model's training on datasets characterized by the specific $\sigma_8$ from \cite{2020A&A...641A...6P}. The model's ability to extrapolate to different variances remains limited. At higher masses, however, the network's predictions correspond more closely with the expected HMF. This partial alignment suggests that the network possesses some degree of generalization capability. Nonetheless, for reliable application across varying cosmologies, incorporating these scenarios into the training set is essential.

\section{Discussion \& Conclusions}\label{sec_Conclusions}

We present a novel approach to understand and predict halo formation from the initial conditions employed in N-body simulations. Benchmark tests indicate that our model can predict Lagrangian FoF-halo regions for simulations efficiently, taking around 7 minutes in a GPU for a simulation with $256^3$ particles in a volume of $50 h^{-1}\mathrm{Mpc}$. For those interested in leveraging or further enhancing our work, we have made our codes publicly available: 
\url{https://github.com/daniellopezcano/instance_halos}.

Our model consists of a semantic network that reliably recognizes regions in Lagrangian space where haloes form, and an instance segmentation network, that identifies individual haloes from the semantic output. Our predictions accurately reproduce simulation results and outperform traditional analytical, semi-analytical techniques, and prior ML methods.

The foundation for our instance segmentation model is the Weinberger approach, first introduced by~\cite{2017arXiv170802551D}. This technique lets us develop a more general framework for identifying Lagrangian halo patches than previous attempts. Employing the Weinberger loss approach, we bypass some limitations of other instance segmentation methods, like the watershed technique employed by~\cite{2020MNRAS.496.5116B}. With our approach, we manage to predict the complicated Lagrangian shapes of haloes that are formed in N-body simulations. This is notably more difficult than the predictions of spherical Peak-Patch-haloes that were considered by~\cite{2019MNRAS.482.2861B}.

Additionally, we quantify in how far halo formation is indetermined by the resolved scales of the initial conditions, to establish an optimal performance limit of machine learning methods. We infer this limit by comparing two simulations which only differ in their initial conditions realization on scales beyond the resolution level. We find an agreement between our model predictions and reference simulations similar to the agreement between the two 'baseline' simulations. This shows that our model extracts information encoded in the initial conditions close to optimal. We suggest that such reference experiments may also be used as a baseline in other ML studies to establish whether information is extracted optimally.

Upon evaluating our semantic model, we measure an accuracy of $0.864$ and an $\textrm{F}_1$-score of $0.838$. Compared to the baseline simulations, which have an accuracy of $0.903$ and an $\textrm{F}_1$-score of $0.884$, our model results stand remarkably close, demonstrating its capability to predict halo regions nearly matching N-body simulations' natural variability.

We also assess our instance segmentation network using various metrics. As depicted in \Fig{fig_7}, our model closely aligns with the baseline across a broad mass range, outperforming previous methods like \cite{2020arXiv201110577L}. We speculate that our approach benefits from the physical constraint that different particles that belong to the same halo are assigned the same halo mass.
Moreover, the halo mass function (HMF) predictions in \Fig{fig_HMFs} closely match the true ground truth values across three orders of magnitude. The visual representations in \Fig{fig_6} reinforce our model's precision, faithfully replicating Lagrangian halo patch positions and shapes.

We have tested through experiments how the network reacts to systematic modifications of the initial conditions. We find that the network correctly captures the response to density perturbations at the finite boxsize provided in the training set. However, it struggles to generalize to larger boxsizes and to cosmologies with different amplitudes of the density field $\sigma_8$. This can easily be improved by increasing the diversity of the training set.

Further, we have found that our network utilizes information from the potential field that is not encoded in the density field of any finite region. Modifications to a large-scale tidal field are consistent with the same linear density field, but do affect the potential landscape. Our network predicts that such tidal fields affect the Lagrangian shape of haloes in an anisotropic manner which is consistent with the intuitive expectation of how a tidal field accelerates and decelerates the infall anisotropically.

We have demonstrated the robustness of our model in its current applications and we believe it could find potential utility in several other scenarios like crafting emulated merger trees, aiding separate-universe style experiments~\citep[e.g.][]{lazeyras_2016, 2021MNRAS.503.1473S} and informing the development of analytical methods for halo formation~\citep[e.g.][]{2021MNRAS.508.3634M, 2023MNRAS.523L...4M}. Other works such as \texttt{MUSCLE-UPS} \citep[][]{2021MNRAS.505.2999T} can also benefit from our semantic predictions alone by informing their algorithm about which particles will collapse into haloes.

Additionally, our model can be used to help understand the development of spin and intrinsic alignments in haloes and galaxies by establishing how tidal fields modify the Lagrangian shapes of haloes. This is a vital ingredient to predict the spin of haloes through tidal torque theory \citep{white_1984}. Also, we can employ our model to predict changes in the Lagrangian regions of halos in combination with the ``splice'' technique presented by \cite{2021MNRAS.508.1189C}. We believe this approach can provide new insights regarding how modifications in the environment of haloes at initial conditions can affect their final properties. We encourage experts in these fields to use our open-source code as a basis for tackling and exploring these and other related problems. 

The models we have presented in this paper can be easily extended to characterize other properties of halos. One possible extension of the model would be to include an additional spatial dimension to our instance network's output to predict final halo concentrations. In this extension of our model, each particle would have associated a concentration prediction whose average (over all particle members of the same halo) would be trained to minimize the mean square error with respect to the true halo concentration.

The findings presented in this work are promising but there exist some aspects of our models that would benefit from further investigation. For instance, extending our methodology to understand other halo properties beyond mass would be a logical next step. It would also be interesting to test our model's performance under a wider variety of simulation conditions, including variations in cosmology and redshift. An additional avenue of exploration might involve delving into capturing intricate structural details, specifically the gap features in the predicted Lagrangian halo regions. Generative Adversarial Networks (GANs) are tools that have demonstrated potential in reproducing data patterns in the context of cosmological simulations~\citep[e.g.][]{2018ComAC...5....4R, 2021ApJ...915...71V, 2021arXiv211106393S, 2022MNRAS.514.3692R, 2023arXiv230805145N, 2023arXiv230512222Z}. Hence, employing a GAN-like approach might help recreate these gap features, further improving our model's ability to mimic the structures of haloes found in N-body simulations.

In conclusion, this study showcases the potential of machine learning for facilitating the study of halo formation processes in the context of cosmological N-body simulations. We provide a fast model that exploits the available information close to optimally. We hope our approach serves as a useful tool for researchers working with N-body simulations, opening avenues for future advancements.

\begin{acknowledgements}
    DLC, JS, MPI and REA acknowledge the support of the ERC-StG number 716151 (BACCO) and of project PID2021-128338NB-I00 from the Spanish Ministry of Science. DLC further thanks Lurdes Ondaro-Mallea for helping to run the baseline simulations, Rodrigo Voivodic for generating the \textsc{ExSHalos} catalogues, and Luisa Lucie-Smith, Drew Jamieson, Francisco Maion, Matteo Zennaro, and Daniel Muñoz-Segovia for their helpful discussions and comments. This work used the following software: \textsc{Python}, \textsc{Matplotlib} \citep{Matplotlib-Hunter07}, \textsc{TensorFlow} \citep{tensorflow2015-whitepaper}, and \textsc{Numpy} \citep{Numpy-vanDerWalt11}.
\end{acknowledgements}

\begin{appendix} 
\section{Watershed segmentation}\label{sec_A1}
In this appendix, we present an alternative approach to instance segmentation, based on the watershed approach. Originally we tried this technique to address the instance segmentation problem, but we finally decided to use the Weinberger approach presented in the main paper because of its theoretical advantages. These are that the loss function closer reflects the objective, that it is possible to predict disconnected regions, and that it is not necessary to define borders. However, during our exploration, we have gained some insights of how to make watershed-based instance segmentation techniques work for friends-of-friends proto-haloes. We will explain these here for the benefit of future studies.

Our watershed approach makes use of a U-Net-based architecture~\cite{2015arXiv150504597R}, specifically a 3D Residual U-Net based on previous work~\cite{franco2021stable}. The model's input consisting of $128\times128\times128\times2$ voxels for ($x$, $y$, $z$, $channels$) axes. The two input \textit{channels} correspond to the initial density field and the potential. 

\begin{figure}
  \resizebox{\hsize}{!}{\includegraphics{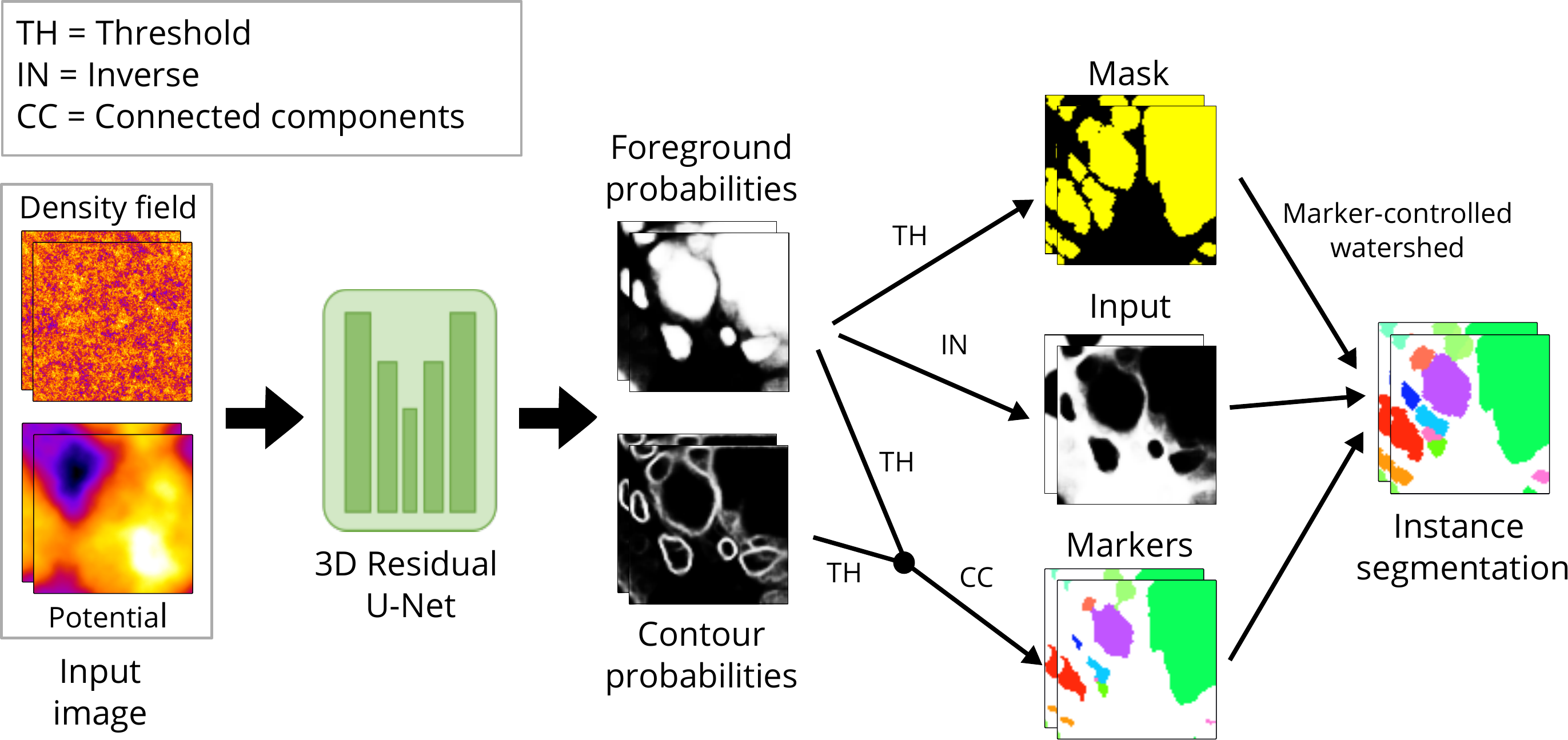}}
  \caption{Processing pipelines of our watershed segmentation approach. The input 3D image contains two channels: the density field and the potential. The model predicts foreground and contour probabilities that are fused to create three inputs for a marker-controlled watershed to produce individual instances.}
  \label{fig:mask_channels}
\end{figure}

The model is trained to predict two output channels: binary foreground segmentation masks and instance contours masks. Following the prediction, the two outputs are thresholded (automatically using Otsu's method~\cite{otsu1979threshold}) and combined. Next, a connected components operation is applied to generate distinct, non-touching halo instance seeds. Subsequently, a marker-controlled watershed algorithm~\cite{meyer1994topographic} is applied, using three key components: 1) the inverted foreground probabilities as the input image (representing the topography to be flooded), 2) the generated instance seeds as the marker image (defining starting points for the flooding process), and 3) a binarized version of the foreground probabilities as the mask image (constraining the extent of object expansion). To binarize the latter, we employed a threshold value of $0.372$, which was determined through the application of the identical methodology outlined in Appendix~\ref{sec_A25}. The collective implementation of these components facilitates the creation of individual halo instances (see Fig.~\ref{fig:mask_channels} for a visual representation). This strategy has been extensively employed within the medical field with remarkable success~\cite{wei2020mitoem,lin2021nucmm,andres2023cartocell}.

\begin{figure}
  \resizebox{\hsize}{!}{\includegraphics{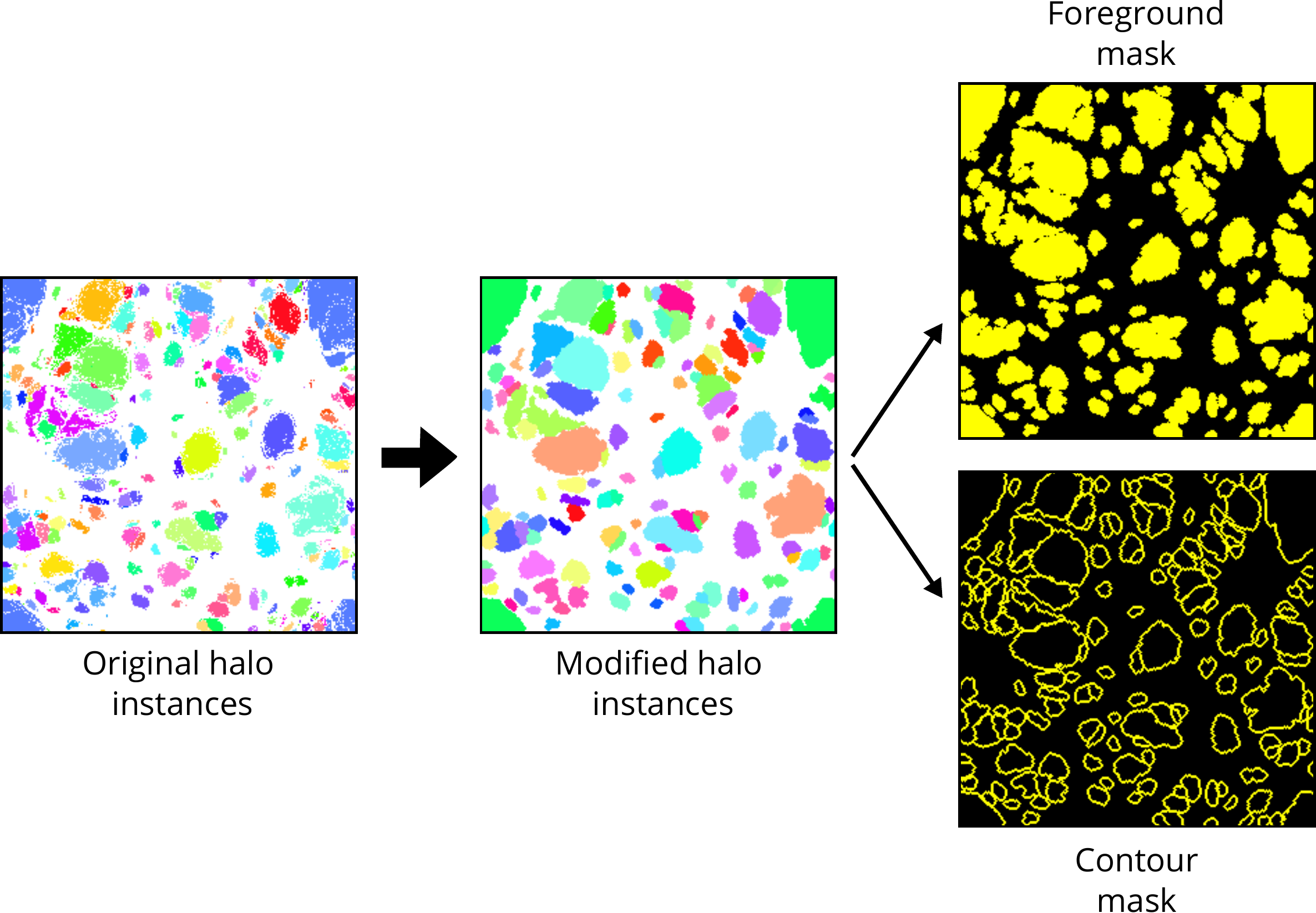}}
  \caption{Data preparation process of our watershed segmentation approach. From left to right: the original halo instances for the considered prediction problem, subsequent modifications involving the removal of small holes and spurious pixels and contour smoothing, and the presentation of both the foreground and contour masks utilized for model training. Pixels coloured in white do not belong to any halo. Pixels with the same colour belong to the same halo and different colours indicate different haloes.}
  \label{fig:data_preparation}
\end{figure}

In order to facilitate the generation of the two channels used to train the network, several transformations were applied to the labels. For each halo instance, small particles along the edges were removed, central holes were filled, and the labels were dilated by one pixel. This process results in instances with smoother boundaries, thereby aiding the network in training (see Fig.~\ref{fig:data_preparation}).

\begin{figure}
  \resizebox{\hsize}{!}{\includegraphics{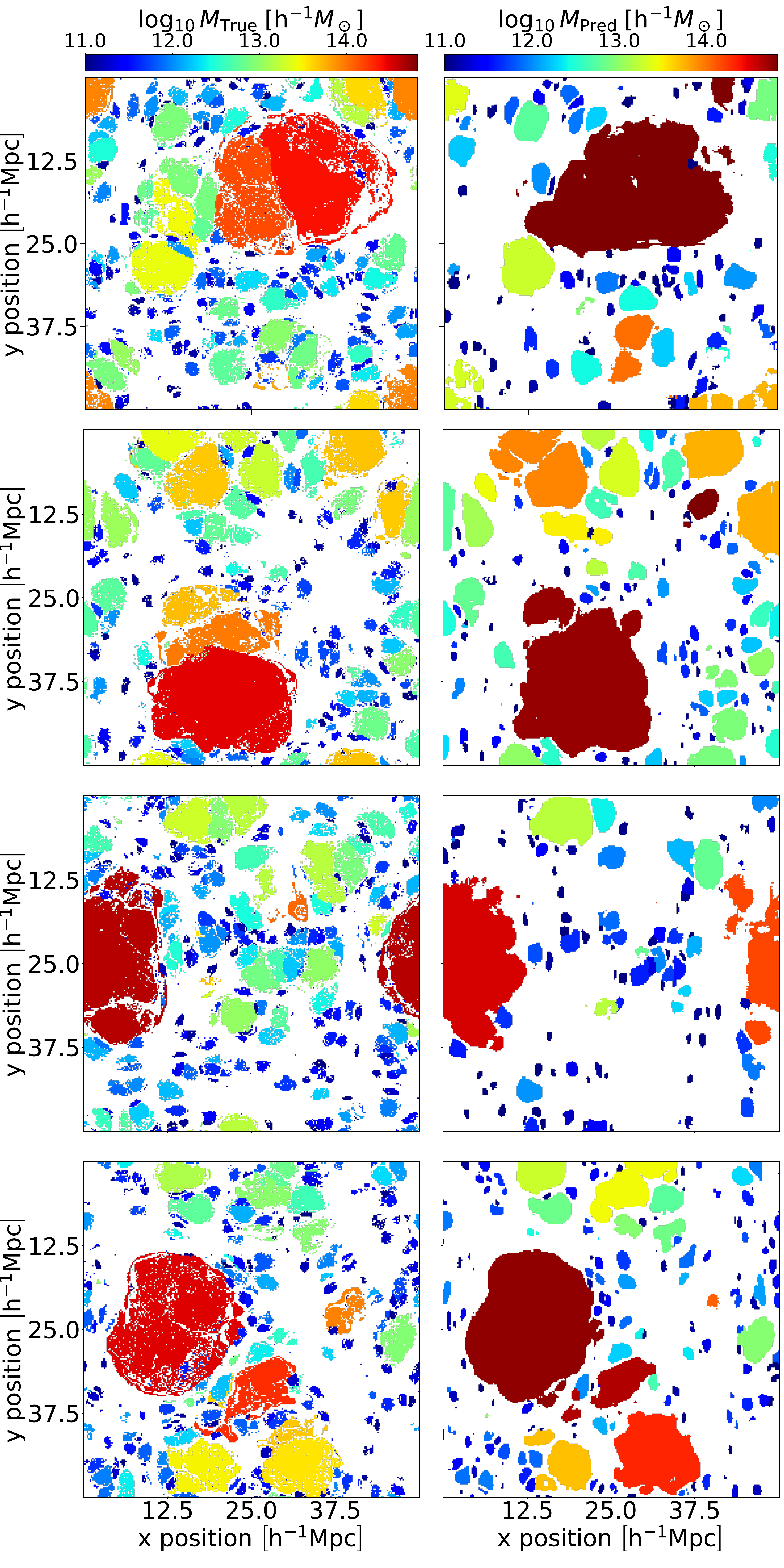}}
  \caption{Results of our watershed segmentation approach presented in an analogous way to results from \Fig{fig_6}.}
  \label{fig_A3}
\end{figure}

The result of this method is depicted in Fig.~\ref{fig_A3}. The code is open source and readily available in BiaPy~\cite{franco-barranco2023biapy}. 

\section{Clustering algorithm}\label{sec_A2}

In this appendix, we describe the clustering algorithm that we have developed. This algorithm calculates instance predictions from the pseudo-space representations that are output by our instance segmentation network.

As described in \Sub{subsec_23_panoptic_segmentation}, the output of our instance network consists of a set of points that populate an abstract space (referred to as pseudo-space). Our instance network has been trained to minimize the Weinberger loss function~\ref{eq_Weinberger}, hence, we expect that the predicted mapping of points in the pseudo-space causes that points corresponding to the same instances to be close to each other, and separated to points that correspond to different instances. In the ideal case where $\mathcal{L}_{\textrm{Wein}}$=0, all points belonging to the same instance would be no farther apart from each other than a distance $2\cdot \delta_{\text{Pull}}$, and the points corresponding to separate instances would be, as close as a distance $2\cdot \delta_{\text{Push}} - \delta_{\text{Pull}}$ close to each other. However, we cannot expect that our network always separates perfectly the different instances. For example, if some Lagrangian voxel has a 60\% chance to belong to halo A and a 40\% chance to belong to halo B, then the optimal location in pseudo space (that statistically minimizes the loss) may be somewhere in between the centre of halo A and B in pseudo space and not inside the $\delta_{\text{Pull}}$ radius of neither. Therefore, we employ a clustering algorithm that can segment the pseudo-space distribution of points also when $\mathcal{L}_{\textrm{Wein}}$ is not exactly zero.

For this, we first estimate the local pseudo-space density $\rho_i$ for each point $i$. For this we compute the distance $r_{k,i}$ to the $k$th-nearest neighbour of the point and assign 
\begin{align}
    \rho_i = \frac{3 k}{4 \pi r_{k,i}^3}
\end{align} 
where $k = N_{\rm{dens}}$ is a hyper-parameter of the clustering algorithm. We accelerate this step with the \textsc{ckd-tree} from the \textsc{scipy} package in \textsc{python}~\citep{2020SciPy-NMeth}. 

Then we determine groups as the descending manifold of the maxima that exceed a persistence ratio threshold $\rho_{\rm{max}} / \rho_{\rm{sad}} \geq p_{\rm{thresh}}$ between maximum and saddle-point. The descending manifold corresponds to the set of particles from whose location following the local density gradient would end up in the same maximum \citep[e.g.][]{disperse,ttk}. For this, we use a slightly modified version of the density segmentation algorithm used in \textsc{subfind} \citep{2001MNRAS.328..726S}: 

We consider the particles from highest to lowest density. For each particle we consider from the $N_{\rm{ngb}}$ nearest particles the subset of particles that have a higher density than $\rho_i$ (this set may be empty). Among these we select the set $B_i$ of the (up to) two closest particles. This set can have zero, one or two particles.

    \begin{itemize}
        \item If the set $B_i$ is empty, then there is a density maximum $\rho_{\rm{max}} = \rho_i$ and we start growing a new subgroup around it.
        \item If the set $B_i$ contains a single particle or two particles that are of the same group, the particle i is attached to the corresponding group.
        \item If $B_i$ contains two particles of different groups, then $i$ is potentially a saddle-point. We check whether the group with the lower density maximum $\rho_{\rm{max}}$ has a sufficient persistence $\rho_{\rm{max}} / \rho_i \leq p_{\rm{thresh}}$. If not, then we merge the two groups (and keep the denser maximum). Otherwise, we keep both groups and we assign the particle to the group of the denser particle in $B_i$. (This step corresponds to following the local discrete density gradient.)
    \end{itemize}
    
Note that unlike the \textsc{subfind} algorithm, we merge groups not at every saddle-point, but only if they are below a persistence threshold. Therefore, sufficiently persistent groups are grown beyond their saddle point and ultimately correspond to the descending manifold of their maximum.

The clustering algorithm has three hyper-parameters $N_{\rm{dens}}$, $N_{\rm{ngb}}$ and $p_{\rm{thresh}}$. We have done a hyper-parameter optimization over these and found that $N_{\rm{dens}} = 20$, $N_{\rm{ngb}} = 15$ (quite close to the default parameters in the \textsc{subfind} algorithm, 20 and 10 respectively) and $p_{\rm{thresh}} = 4.2$ give the best results, though our results are not very sensitive to moderate deviations from this. We can understand the quantitative value of the persistence ratio threshold by considering that the relative variance of our density estimate is
\begin{align}
    \sigma_{\log \rho} \approx \frac{\sigma_\rho}{\rho} = \frac{1}{\sqrt{N_{\mathrm{dens}}}} \approx 0.22
\end{align}
so that at a fixed background density having a density contrast of $p_{\rm{thresh}} = 4.2$ due to Poisson noise corresponds to a 
\begin{align}
    \Delta \log \rho = \log(p_{\rm{thresh}}) \approx 1.43 \approx 6.5 \sigma_{\log \rho}
\end{align}
outlier. Therefore, the persistence ratio threshold $p_{\rm{thresh}}$ ensures that it is very unlikely that our algorithm mistakes a spurious overdensity in the pseudo space for a group.

\section{Semantic threshold}\label{sec_A25}

\begin{figure}
  \resizebox{\hsize}{!}{\includegraphics{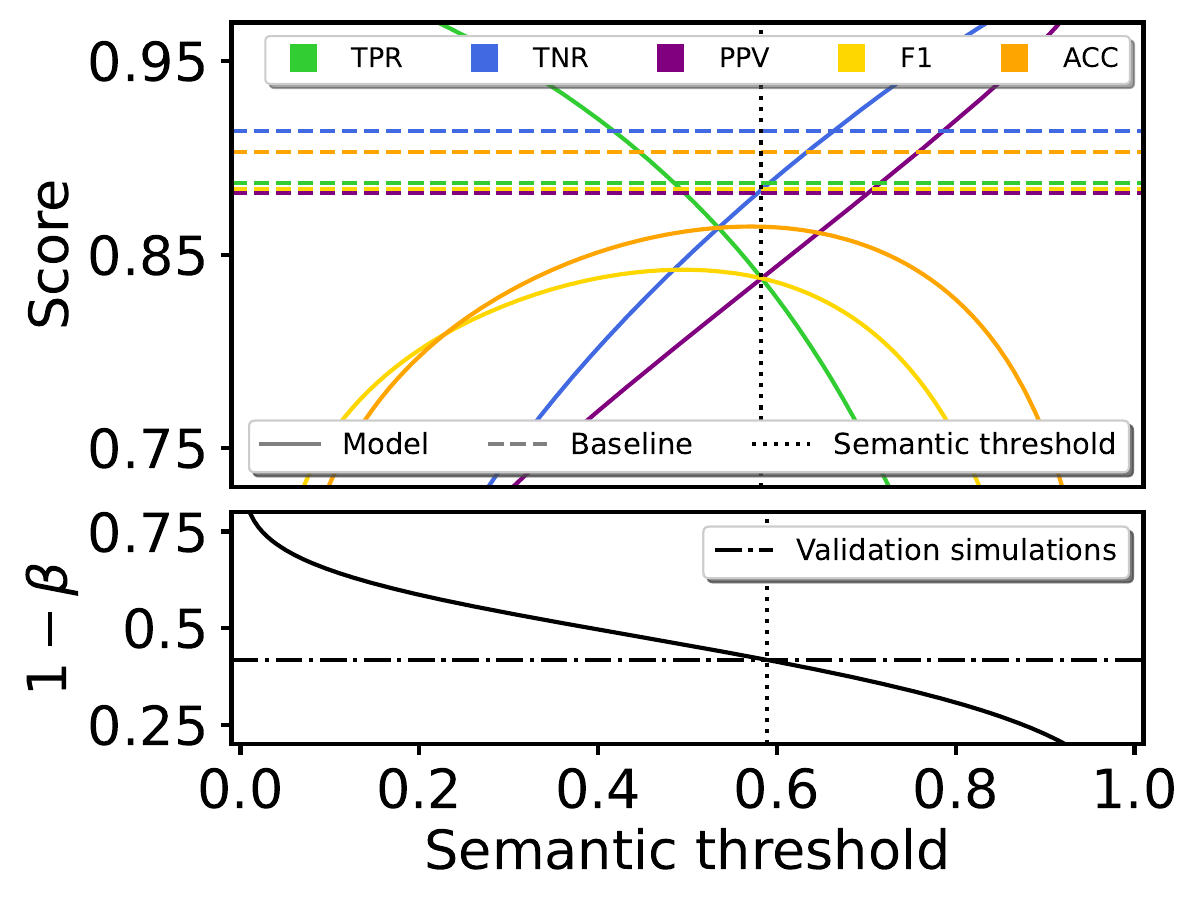}}
  \caption{Top panel: Evolution of different metrics (TPR - green, TNR - blue, PPV - purple, $\textrm{F}_1$-score - yellow \& ACC - orange) measured employing the predictions of the semantic model as a function of the semantic threshold selected (solid lines); we also show the values measured for the corresponding metrics studying the differences between the baseline simulations (horizontal dashed lines). Bottom panel: Fraction of voxels predicted to be collapsed (equivalent to $1 -\beta$) as a function of the semantic threshold employed (solid black line); the horizontal black dashed line corresponds to the fraction of particles that end up in DM haloes measured in the validation simulations. In both panels, the vertical black dotted line shows the semantic threshold we employ; this threshold has been selected to match the fraction of collapsed voxels.}
  \label{fig_semantic_metrics}
\end{figure}

In the bottom panel of \Fig{fig_semantic_metrics} we present how the predicted fraction of voxels that are members of a halo (that is $1-\beta$) evolves as we change the semantic threshold (black solid line). As it can be expected when the semantic threshold is close to zero, the majority of voxels are identified as members of haloes, and the contrary occurs when the semantic threshold approximates one. The horizontal dashed-dotted line corresponds to the ground truth value of $1 -\beta = 0.418$, measured in the validation simulations. The semantic threshold value that we have selected is $0.589$ (black dotted vertical line). This value corresponds to the intersection between the black solid line and the dashed-dotted line; it ensures that the total fraction of voxels that are members of haloes is correctly reproduced. Choosing this criterion to determine the semantic threshold also ensures more robust instance predictions since the number of FP cases is reduced, hence eliminating potentially uncertain pseudo-space particles that would complicate the clustering procedure.

In the top panel of \Fig{fig_semantic_metrics} we show the evolution of several metrics as a function of the semantic threshold value. These metrics allow us to asses the quality of our semantic predictions by comparing our results with values obtained using the baseline simulations. We study the behaviour of five different metrics: True Positive Rate $\mathrm{TPR}$, True Negative Rate $\mathrm{TNR}$, Positive Predictive Value $\mathrm{PPV}$, Accuracy $\mathrm{ACC}$ and the $\textrm{F}_1$-score.

In the top panel of \Fig{fig_semantic_metrics} we also present the values obtained for the different metrics using the baseline simulations (horizontal dashed lines). We have obtained these results considering one of the baseline simulations as predicted maps and the other simulation as the ground truth. The values measured for the different metrics in the baseline simulations give us an expected ideal performance that we would like to reproduce with our model.

If we focus on the performance curves for the accuracy and the $\textrm{F}_1$-score (orange and yellow lines respectively) we can appreciate that they always remain under the baseline limit. The curve for the $\textrm{F}_1$-score peaks around the value for the semantic threshold of $0.5$,  which is a behaviour we expected since we considered the balanced cross-entropy loss to train our semantic model. The value for the $\textrm{F}_1$-score at its maximum is $\textrm{F}_1(0.5) = 0.842$, which is very similar to the value at the point in which we have fixed the semantic threshold, $\textrm{F}_1(0.589) = 0.838$. The $\textrm{F}_1$-score obtained is only about $5\%$ away from the optimal value obtained from the baseline simulations $\textrm{F}_1^{\mathrm{Chaos}} = 0.884$. The accuracy reaches its maximum value around the semantic threshold of $0.58$, where $\mathrm{ACC}(0.58) = 0.864$; the value for the model accuracy is even closer to the baseline limit $\mathrm{ACC}^{\mathrm{Chaos}} = 0.903$.

\section{Generate full-box predictions from crops}\label{sec_A3}

In this appendix, we address the challenge of generating full-box predictions employing our instance segmentation model.

While our network architecture captures intricate features within simulation sub-volumes, the challenge arises when we aim to apply it to arbitrarily large input domains. Unlike some other ML approaches that rely on networks that are translational invariant, our model incorporates the Lagrangian positions of particles as input channels, making it dependent on the relative Lagrangian position. This design choice ensures that similar regions of the initial density field are mapped to distinct locations in the pseudo-space, allowing us to distinguish between separate structures, even if they are locally identical. However, this feature also presents a challenge when creating full-box predictions. Combining independent crop predictions straightforwardly may lead to inconsistencies due to the network's inherent non-translational invariance. To tackle this issue, we have developed a methodology for predicting sub-volumes independently and then merging these predictions to generate accurate full-box instance segmentation results.

To reduce the boundary effects that may result from such a method we employ the following strategy.
\begin{enumerate}

    \item \label{item:step1} We evaluate the instance network centred several times, centred on locations $\Vec{q}_{ijk}$ that are arranged on a grid
    \begin{align}
        \Vec{q}_{ijk} &= \begin{pmatrix}i \cdot n_{\rm{off}} \\ j \cdot n_{\rm{off}} \\ k \cdot n_{\rm{off}} \end{pmatrix},
    \end{align}
    where we choose an offset of $n_{\rm{off}} = 64$ voxels and $(i,j,k)$ run so far that the whole periodic volume is covered -- e.g. from 0 to 4 each for a $256^3$ simulation box. The network's input in each case corresponds to the $144^3$ voxels (periodically) centred on $\Vec{q}_{ijk}$ and the instance segmentation output will predict labels for the $128^3$ central voxels.
    
    \item \label{item:step2} From each prediction we only use the predicted labels of the central $n_{\rm{off}}^3 = 64^3$ voxels, since we expect these to be relatively robust to field-of-view effects. We combine these from all the predictions to a global grid that has the same dimensions as the input domain. In this step we add offsets to the labels so that the labels that originate from each predicted domain are unique in the global grid (this process will become relevant in step \ref{item:step4} where we define a graph used to link instances).
    
    \item We repeat steps \ref{item:step1}-\ref{item:step2}, but with an additional offset of $(n_{\rm{off}}/2, n_{\rm{off}}/2, n_{\rm{off}}/2)^T$. We additionally offset the labels in this second grid so that no label appears in both grids.
    
    \item \label{item:step4} We use the two lattices and the intersections between instances to identify which labels should correspond to the same object. We do this by creating a graph\footnote{using the \textsc{networkx} library \citep{hagberg2008exploring}} where each instance label is a node. Initially the graph has no edges, but we subsequently add edges if two labels should be identified (i.e. correspond to the same halo). Each connected component of the graph will then correspond to a single final label. To define the edges of the graph, we consider each quadrant $Q$ of size $(n_{\rm{off}}/2)^3$ individually, since such quadrants are the maximal volumes over which two labels can intersect. We define the intersection $I_Q(l_1,l_2)$ of two labels $l_1$ and $l_2$ as the number of voxels that both carry label $l_1$ in grid one and label $l_2$ in grid two. We define as the union $U_Q(l_1,l_2)$ the number of voxels inside of quadrant $Q$ that carry $l_1$ in grid 1 or $l_2$ in grid 2 (or both). We then add an edge  between $l_1$ and $l_2$ into the graph if for any quadrant $Q$ it is
    \begin{align}
        \frac{I_Q(l_1,l_2)}{U_Q(l_1,l_2)} \geq IoU_{\rm{thresh}}
    \end{align}
    where we set $IoU_{\rm{thresh}} = 0.5$.
    
    \item We summarize each connected component in the graph into a new label. After this operation for most voxels the new label in grid 1 and in grid 2 agree and we can choose that label as our final label. However, for a small fraction of voxels the labels still disagree, because the corresponding instances had too little overlap to be identified with each other. In this case, we assign to the corresponding voxel the label that contains the larger number of voxels in total.
    
\end{enumerate}

\begin{figure}
\begin{center}
  \resizebox{0.9\hsize}{!}{\includegraphics{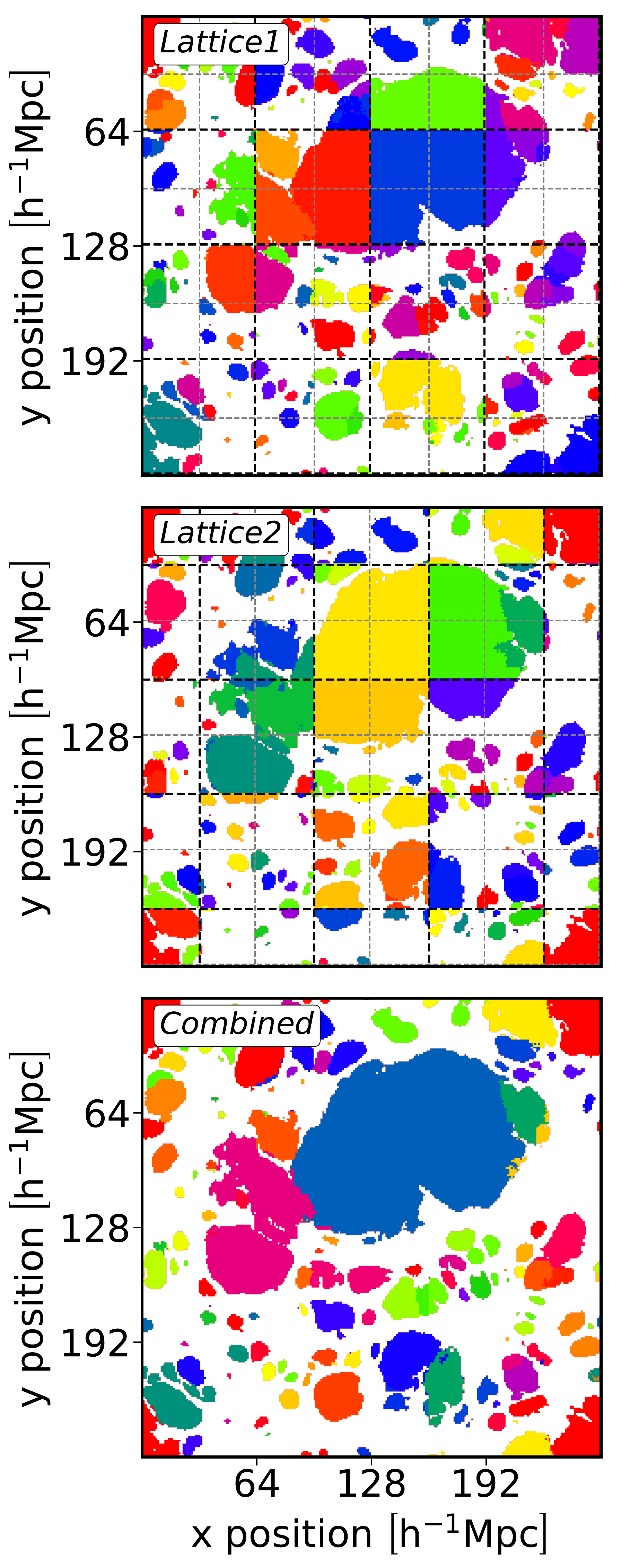}}
  \caption{Process of merging predictions from two overlapping lattice structures to produce a full-box instance segmentation map. 'Lattice1' (\textbf{top}) and 'Lattice2' (\textbf{middle}) represent predictions from initial and shifted lattice grids, respectively, with unique color-coded labels for instances. Black dashed lines indicate the lattice employed in each case, while thin dashed grey lines correspond to the lattice employed in the reciprocal scenario. 'Combined' (\textbf{bottom}) depicts the final synthesized full-box map, where instances have been merged based on their overlap, demonstrating the effectiveness of the methodology in generating contiguous and comprehensive halo segmentations from smaller, predicted sub-volumes.}
  \label{fig_appD}
\end{center}
\end{figure}

We illustrate the different steps of this procedure in \Fig{fig_appD}. The top panel, labelled 'Lattice1', shows the individual instances predicted in the first lattice arrangement. Each colour represents a distinct label assigned to a group of voxels within the $64^3$ central region of the sub-volumes. The middle panel, 'Lattice2', displays the second set of predictions using a shifted lattice by half the offset in each dimension. Here again, different colours represent unique instance labels. The bottom panel, 'Combined', presents the final merged full-box prediction. It is generated by synthesizing the labels from 'Lattice1' and 'Lattice2' using the graph-based method to connect overlapping instances. The resulting image shows larger, coherent structures, indicative of the correct performance of combining both lattices.

Regarding the semantic segmentation network, we can merge the predictions corresponding to different crops independently since, in this case, we are truly working with a translation-invariant network. We employ the central $64^3$ voxels (analogous to 'Lattice1') of separate predictions and merge them together to generate the final full-box predictions of the semantic segmentation network.

\section{Comparison with ExSHalos}\label{sec_A4}

In this appendix, we explore how the results obtained with the \textsc{ExSHalos} code~\citep{2019arXiv190606630V} compare against our semantic and instance predictions.

As mentioned in \Sub{subsec_31Semantic}, \textsc{ExSHalos} is an explicit implementation of the excursion set theory that identifies haloes in Lagrangian space by growing spheres around density peaks until the average density inside crosses a specified barrier for the first time. The barrier shape is motivated by the ellipsoidal collapse~\citep{2001MNRAS.323....1S, 2011MNRAS.418.2403D} and we have fitted the three free parameters in the model to reproduce the mean halo mass function of our simulations.

\begin{figure*}
\begin{center}
\includegraphics[width=1.\textwidth]{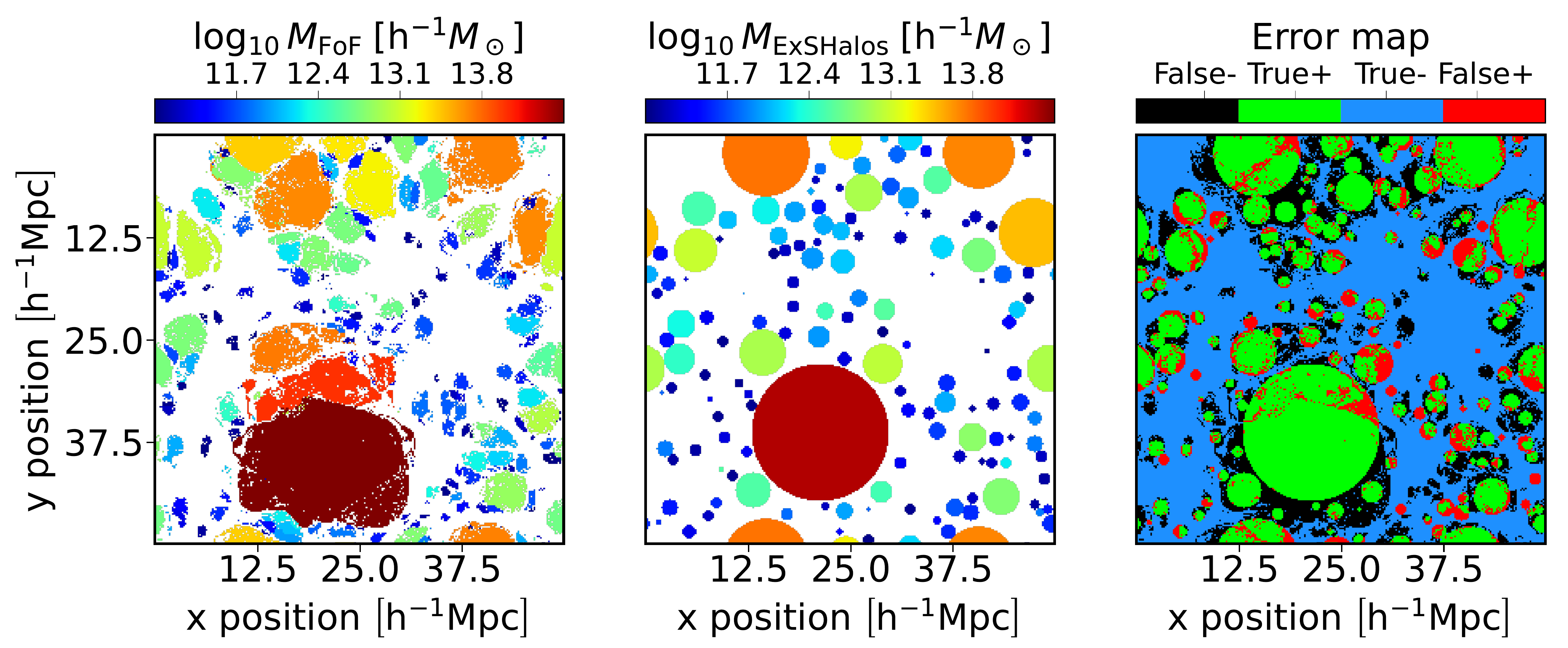}\vspace*{0.cm}
\caption{Slices through the Lagrangian field of friends-of-friends proto-haloes, and the corresponding predictions using the \textsc{ExSHalos} algorithm. \textbf{Left panel}: ground truth masses obtained using N-body simulations (friends-of-friends proto-haloes). \textbf{Central panel}: predicted masses obtained using the \textsc{ExSHalos} algorithm. \textbf{Right panel} (analogous to left panel of \Fig{fig_4}): Semantic pixel-level error map between \textsc{ExSHalos} and friends-of-friends haloes indicating true positive (green), true negative (blue), false negative (black), and false positive (red) regions.
}
\label{fig_ExSHalos_map}
\end{center}
\end{figure*}

In \Fig{fig_ExSHalos_map} we show a map-level comparison between the Lagrangian proto-haloes identified in one of our validation simulations with the friends-of-friends algorithm (left panel), and the \textsc{ExSHalos} detected employing the code presented in~\cite{2019arXiv190606630V} (central panel). The \textsc{ExSHalos} regions in Lagrangian space are spherical by construction (see the middle panel of \Fig{fig_ExSHalos_map}). The physical approach of the \textsc{ExSHalos} algorithm enables to identify, with a reasonable degree of accuracy, the location of proto-haloes in Lagrangian space, and their mass. However, the built-in assumption that proto-haloes are spherical gives only a crude approximation to the actual proto-halo shapes. In Table~\ref{tab_confusion_matrix} we quantify the differences between \textsc{ExSHalos} and friends-of-friends employing several semantic metrics.

\begin{figure}
  \resizebox{\hsize}{!}{\includegraphics{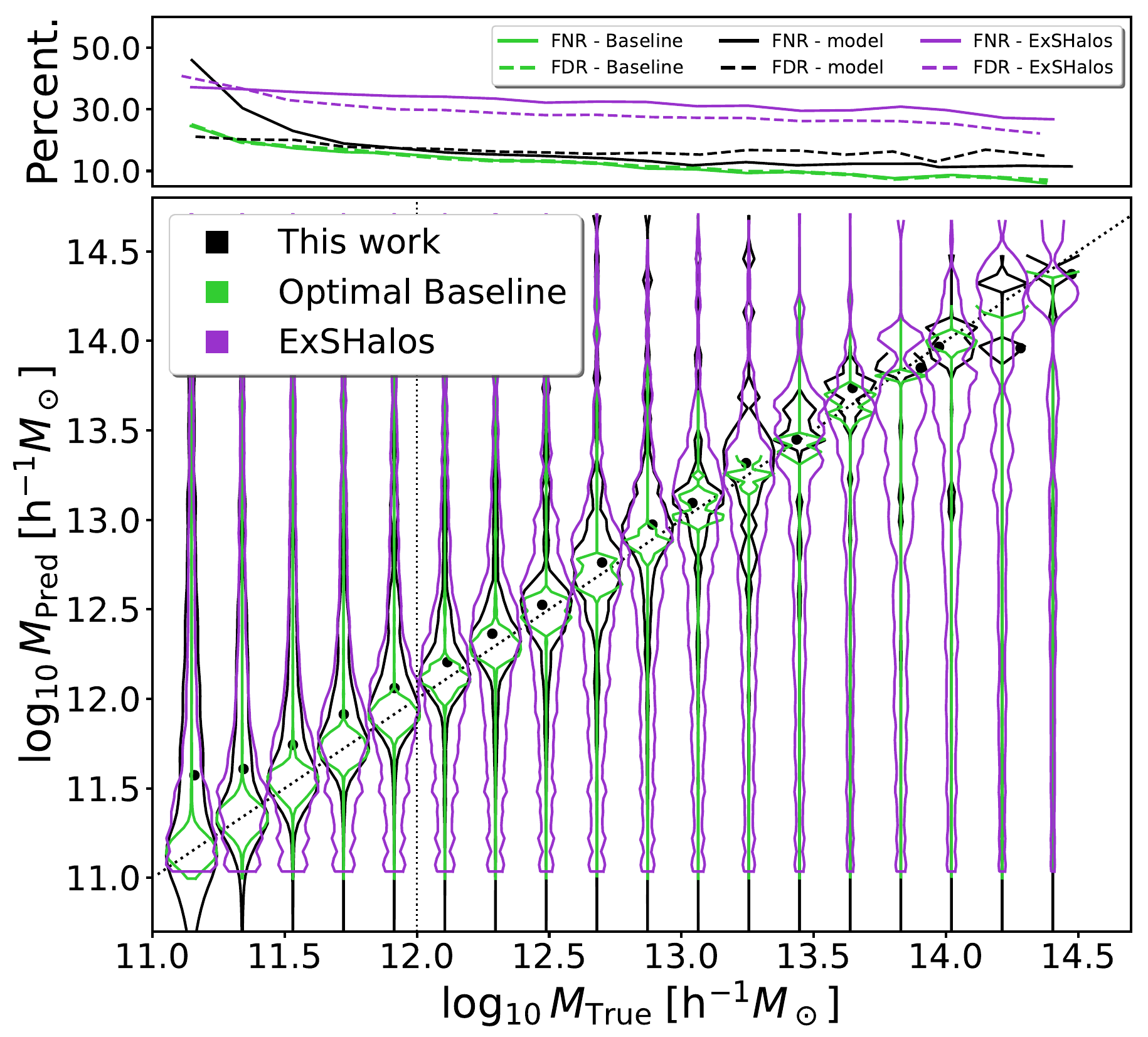}}
  \caption{``Violin plot'', visualizing the distribution of predicted halo masses (at a voxel level) for different ground-truth mass bins. The black violin plots show the results obtained with our instance segmentation model. Green violin plots show the agreement between the two baseline simulations -- representing an optimal target accuracy. The purple violin plots in the main panel correspond to the comparison with the \textsc{ExSHalos} predictions. The solid black line in the top panel shows the false negative rate, FNR, as a function of the ground truth halo mass. The dashed black line represents the fraction of predicted collapsed pixels that are not collapsed as a function of predicted halo mass (false discovery rate, $\textrm{FDR}$). The green and purple lines on the top panel correspond to the analogous results obtained from the baseline simulations and \textsc{ExSHalos} respectively.
  }
  \label{fig_ExSHalos_violin}
\end{figure}

In \Fig{fig_ExSHalos_violin} we present a violin plot analogous to \Fig{fig_7}. This plot shows a comparison between the ground truth halo masses (friends-off-friends) and the predicted masses from our model associated with the particles/voxels in our validation set (black violin lines in the main panel). We also include the comparison between the masses of \textsc{ExSHalos} and of friends-off-friends haloes (purple violin lines). We have generated the violin lines of \textsc{ExSHalos} employing all our simulations (both training and validation) to achieve better statistics. Our model predictions are capable of achieving greater mass accuracy than \textsc{ExSHalos} throughout all mass bins considered here.

In the upper panel of \Fig{fig_ExSHalos_violin}, we show the False Negative Rate (FNR) as solid lines against the ground truth halo mass, and the False Discovery Rate (FDR) as dashed lines against the predicted mass. This plot is analogous to the top plot in \Fig{fig_7} (See \Sub{subsec_32Instance} for details). We additionally include solid and dashed purple lines corresponding to the \textsc{ExSHalos} case. It's clear that \textsc{ExSHalos} predicts higher FNR and FDR values compared to the baseline case and our model predictions, indicating more semantically-misclassified particles.

\end{appendix}

% for the bibliography, at the end
\bibliographystyle{aa} % style aa.bst
\bibliography{archive} % your references archive.bib
\end{document}